\documentclass[acmsmall,screen,nonacm]{acmart}

\AtBeginDocument{%
  }
    
\usepackage{enumitem}
\usepackage{graphicx}
\usepackage{makecell}
\usepackage{xcolor}
\usepackage{xspace}
\usepackage{array}
\usepackage{wrapfig} 
\usepackage{colortbl}
\usepackage{booktabs}
\usepackage{arydshln}
\usepackage{hyperref}
\usepackage{soul}
\usepackage{tabularray}
\usepackage{multirow}
\usepackage{tcolorbox}
\usepackage[normalem]{ulem}
\usepackage{pdfpages}

\usepackage{algorithm}
\usepackage{algpseudocode}
\usepackage{amsmath}
\algrenewcommand\algorithmicrequire{\textbf{Input:}}
\algrenewcommand\algorithmicensure{\textbf{Output:}}
\newcommand{\BlueComment}[1]{\hfill\textcolor{blue}{\textit{// #1}}}

\usepackage{subcaption} % in preamble

\newcommand{\tool}[0]{\textit{TokenRepair}\xspace}

\usepackage{hyperref}
\usepackage{url}

\begin{document}

\title{Enhancing Automated Program Repair via Faulty Token Localization and Quality-Aware Patch Refinement}

\author{JIAOLONG KONG}
\affiliation{%
  \institution{Singapore Management University}
  % \streetaddress{P.O. Box 1212}
  \country{Singapore}
}
% \email{jlkong@smu.edu.sg}

\author{XIAOFEI XIE}
\affiliation{%
  \institution{Singapore Management University}
  % \city{Rocquencourt}
  \country{Singapore}
}
% \email{xfxie@smu.edu.sg}

\author{YIHENG XIONG}
\affiliation{%
  \institution{Singapore Management University}
  % \streetaddress{}
  % \city{Hekla}
  \country{Singapore}}
% \email{snowbirds.mf@gmail.com}

\author{YUEKUN WANG}
\affiliation{%
  \institution{Singapore Management University}
  % \streetaddress{}
  % \city{Hekla}
  \country{Singapore}}
% \email{liu.shangqing@ntu.edu.sg}

\author{JIAN WANG}
\affiliation{%
  \institution{Singapore Management University}
  % \streetaddress{}
  % \city{Hekla}
  \country{Singapore}}

\renewcommand{\shortauthors}{Kong et al.}

\begin{abstract}
Large language models (LLMs) have recently demonstrated strong potential for automated program repair (APR). However, existing LLM-based techniques primarily rely on coarse-grained external feedback (e.g., test results) to guide iterative patch generation, while lacking fine-grained internal signals that reveal why a patch fails or which parts of the generated code are likely incorrect. This limitation often leads to inefficient refinement, error propagation, and suboptimal repair performance.
In this work, we propose TokenRepair, a novel two-level refinement framework that enhances APR by integrating internal reflection for localizing potentially faulty tokens with external feedback for quality-aware patch refinement. Specifically, TokenRepair first performs internal reflection by analyzing context-aware token-level uncertainty fluctuations to identify suspicious or low-confidence tokens within a patch. It then applies Chain-of-Thought–guided rewriting to refine only these localized tokens, enabling targeted and fine-grained correction. To further stabilize the iterative repair loop, TokenRepair incorporates a quality-aware external feedback mechanism that evaluates patch quality and filters out low-quality candidates before refinement.
Experimental results show that TokenRepair achieves new state-of-the-art repair performance, correctly fixing 88 bugs on Defects4J 1.2 and 139 bugs on HumanEval-Java, demonstrating substantial improvements ranging from 8.2\% to 34.9\% across all models on Defects4J 1.2 and from 3.3\% to 16.1\% on HumanEval-Java.

% Automated Program Repair (APR) aims to automatically generate patches for rectifying software bugs. Recent advances in large language models (LLMs) have yielded promising results in APR, particularly through conversation-based repair frameworks. Nevertheless, existing conversation-driven APR methods suffer from two critical limitations: 1) \textit{Repair deviates away from the real faulty position}. The LLMs cannot identify the actual faulty tokens, and always repair the irrelevant tokens among the patch instead. 2) \textit{Error propagation across patches}. During multi-turn repair, error in subsequent patches are still same as that in predecessor ones. In this paper, we propose \tool, a novel uncertainty-guided APR approach that leverages token-level uncertainty to do fine-grained fault localization and patch quality measurement, which solves the existing limitations and enables LLMs to explore patch space more effectively and efficiently. We evaluate \tool on two widely-used benchmarks, Defects4J 1.2 and HumanEval-Java, across five prominent large language models. The results demonstrate that \tool significantly outperforms existing methods, achieving a new state-of-the-art in automated program repair. For instance, our method consistently surpasses all baseline methods on Defects4J 1.2, yielding performance improvements ranging from 8.2\% to 34.9\%.

% repairs 88 bugs on Defects4J 1.2 and 139 bugs on HumanEval-Java, while the best-performing baseline only fixes 82 and 131 bugs, indicating the improvements of 7.3\% and 6.1\% respectively.
\end{abstract}

\maketitle

\section{Introduction}
As software systems grow increasingly complex, the emergence of bugs and vulnerabilities has become unavoidable. These defects can precipitate system failures, security breaches, and degraded user experiences. The manual identification and remediation of such issues remains a resource-intensive undertaking, imposing substantial burdens on development teams. Industry reports indicate that annual expenditures on bug detection and remediation exceed billions of dollars~\cite{britton2013reversible}, with developers allocating approximately 50\% of their time to debugging and error correction activities~\cite{latoza2006maintaining, britton2012quantify,zhang2023survey}. In response to these challenges, Automatic Program Repair (APR) has emerged as a promising paradigm, offering automated patch generation to address software defects.

With the rapid advancement of large language models (LLMs), recent research has shown that they are highly effective for automated program repair. Trained on massive corpora of natural language and source code, LLMs exhibit strong capabilities in understanding program semantics, reasoning about code behavior, and generating meaningful edits. Recent studies~\cite{10.1145/3540250.3549101, xia2023automated, prenner2022can, jiang2023impact, sobania2023analysis} consistently demonstrate that LLM-based repair approaches can surpass traditional template-based and conventional deep learning–based APR techniques.

The existing LLM-based APR frameworks mainly follow a multi-iteration patch-refinement paradigm: an LLM first proposes an initial patch, the patch is validated through compilation or test execution, and then the LLM refines the patch based on the feedback~\cite{xia2023automated,10.1145/3719345}. This iterative conversational loop has proven effective for progressively guiding LLMs toward producing correct patches, as each round of feedback has the potential to reduce ambiguity, tighten constraints, and align the model’s reasoning with program execution semantics.

However, existing conversational, multi-iterative APR paradigms primarily rely on {external feedback}, typically the pass/fail outcomes from test execution, to guide subsequent patch generation. While effective, this feedback mechanism has inherent limitations. First, test results are often sparse and coarse-grained: they indicate whether a patch succeeds or fails, but provide little insight into why it failed or which specific code region in this patch is responsible for the failure. This lack of fine-grained diagnostic information leaves a gap in understanding the true root cause of the failure.
Second, because LLMs receive no explicit guidance on the faulty location or underlying defect mechanism, they often refine patches in an uninformed or misguided direction. This may lead the model to repeatedly modify incorrect patches, propagate spurious edits, or drift further away from the correct fix across iterations. As a result, the repair loop can become unstable or inefficient, especially in programs with subtle faults or complex semantics.

\begin{figure}
    \centering
    \includegraphics[width=1\linewidth]{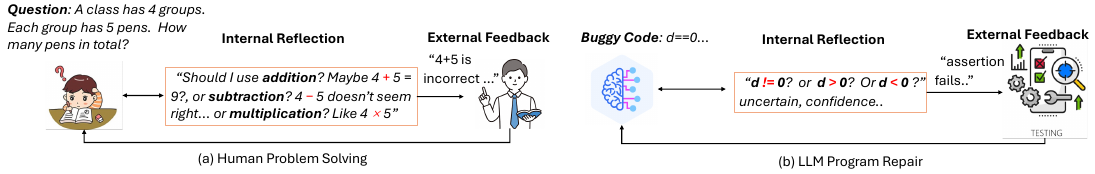}
    \caption{Motivating example of human problem solving}
    \label{fig:placeholder}
\end{figure}

Let us draw inspiration from how humans typically solve problems. Consider a person attempting a math puzzle or logical reasoning task, as shown in Fig.~\ref{fig:placeholder} (a).
Before committing to an answer, they perform \textbf{internal reflection}: they carefully examine the problem and their preliminary solution, assess which steps they feel confident about, and identify portions where their understanding is uncertain. They then reason more deeply about these uncertain components, exploring multiple tentative possibilities or alternative solution paths. This introspective process naturally highlights areas where errors are likely to occur and directs attention toward ambiguous or conceptually challenging parts of the task. In parallel, humans also rely on \textbf{external feedback}. After forming an initial answer, they may consult an answer key, verify with a peer, or check whether the solution satisfies known rules. This feedback provides objective signals about correctness, helping them revise or refine their reasoning accordingly.
Effective problem solving emerges from the interaction of these two complementary signals: internal reflection offers fine-grained, self-generated clues about potential weaknesses, while external feedback delivers objective confirmation that guides refinement toward the correct solution.

However, existing LLM-based APR techniques rely almost exclusively on external feedback, without internal reflection, particularly at the fine-grained level of identifying {which specific tokens in a generated patch may be incorrect}. This introspective ability is crucial for effective repair: understanding which parts of the generated patch are uncertain or potentially faulty would enable fine-grained, token-level refinement, rather than repeatedly regenerating entire patches in a coarse-grained manner, as illustrated in Fig.~\ref{fig:placeholder} (b). 

We argue that effective LLM-based repair should integrate both internal reflection and external feedback, though doing so introduces two main technical challenges. For the internal reflect, \textit{it is challenging to reveal which individual tokens are suspicious within a patch}.  Although recent studies~\cite{spiess2024calibration, zhu2025uncertainty, fu2025deep} show that LLM uncertainty is a useful indicator of generation quality, current uncertainty-based approaches are primarily designed for coarse-grained assessments, typically evaluating the confidence of an entire prediction or local sequence (e.g., a whole execution trace). As a result, they cannot support precise, token-level repair guidance.
On the external feedback side, existing methods typically follow a loop of repeatedly regenerating patches based on test results feedback until a preset budget is exhausted. \textit{The challenges is that how to evaluate whether the current patch is a promising candidate before proceeding to the next iteration}.  Without assessing the quality of the current patch, the repair loop can continue compounding errors, leading the LLM further away from the correct fix and exacerbating error propagation across iterations.

In this work, we propose a novel repair method, named \tool, which performs patch refinement at two complementary levels: token-level refinement guided by internal reflection and patch-level refinement driven by external feedback. To realize this design, we address the two aforementioned challenges.
First, we introduce a context-aware uncertainty–based localization method that identifies suspicious tokens within a generated patch. We design a metric that captures the relative uncertainty fluctuation at each token, measuring how much a token’s uncertainty deviates from that of its predecessor, to derive fine-grained token-level suspiciousness scores. Once these suspicious tokens are localized, we apply Chain-of-Thought (CoT)–guided decoding to selectively refine the erroneous tokens and explore a broader patch space focused on the identified problematic positions.
Second, to mitigate error propagation across iterations during the external-feedback refinement process, we incorporate uncertainty-guided trace quality assessment. In particular, we use the uncertainty of the initial token as a proxy for the overall quality of a generated patch. This design is motivated by prior findings~\cite{zhu2025uncertainty}, which show that the first token typically exhibits the highest prediction difficulty and can thus serve as an informative indicator of the reliability of the entire generation. Patches with high uncertainty are filtered out early, preventing low-quality candidates from misleading the refinement process.

We evaluate \tool on two widely used benchmarks, Defects4J 1.2 and HumanEval-Java, using five prominent large language models. Our evaluation proceeds in three stages. First, we show that the proposed token-level internal reflection mechanism is accurate in localizing faulty tokens, which can achieve averaged accuracy ranging from 0.589 to 0.695, providing the fine-grained guidance necessary for effective patch refinement. Next, we demonstrate that \tool substantially outperforms existing APR approaches, achieving new state-of-the-art results. Specifically, our method correctly repairs 88 bugs on Defects4J 1.2 and 139 bugs on HumanEval-Java, surpassing the best baseline and demonstrating substantial improvements ranging from 8.2\% to 34.9\% across all models on Defects4J 1.2 and from 3.3\% to 16.1\% on HumanEval-Java. Finally, our ablation study confirms the complementary effectiveness of both components: the internal-reflection–guided token refinement and the uncertainty-guided patch quality assessment within the external feedback loop. Without these components, repair effectiveness will decrease by at most 20.6\% on Defects4J 1.2 and 12.0\% on HumanEval-Java. 

In summary, this paper makes the following contributions:
\begin{itemize}[leftmargin=*]
\item 
We are the first to incorporate internal reflection into the LLM-based repair process. By analyzing token-level uncertainty fluctuations, our method identifies faulty or low-confidence tokens within generated patches, enabling fine-grained and targeted refinement rather than coarse-grained patch regeneration.
% We propose an uncertainty-guided APR method for conversation-based APR framework. Specifically, \tool leverages uncertainty metrics derived from model inference 
% to conduct dual-level quality assessment: token-level evaluation for identifying and refining erroneous tokens, and trace-level evaluation for guiding patch selection and repair strategies.

\item 
% We present a novel approach for fine-grained token-level fault localization that incorporates both uncertainty values and fluctuation magnitudes, thereby establishing a robust foundation for patch refinement and facilitating the generation of higher-quality fixes.
We propose \tool, a new APR framework that integrates token-level refinement and patch-level refinement. This joint design improves the repair effectiveness, mitigates error propagation and stabilizes the iterative repair loop.

\item
We perform a comprehensive evaluation to evaluate the effectiveness of our tool. The results demonstrate that \tool achieves new state-of-the-art performance in terms of correct bug fixes, yielding substantial improvements ranging from 8.2\% to 34.9\% across all models on Defects4J 1.2 and from 3.3\% to 16.1\% on HumanEval-Java.
% All experimental data and source code are publicly available ~\cite{ourweb}.
\end{itemize}

\section{Motivation Example}
\begin{figure}[h]
    \centering
    \includegraphics[width=0.9\linewidth]{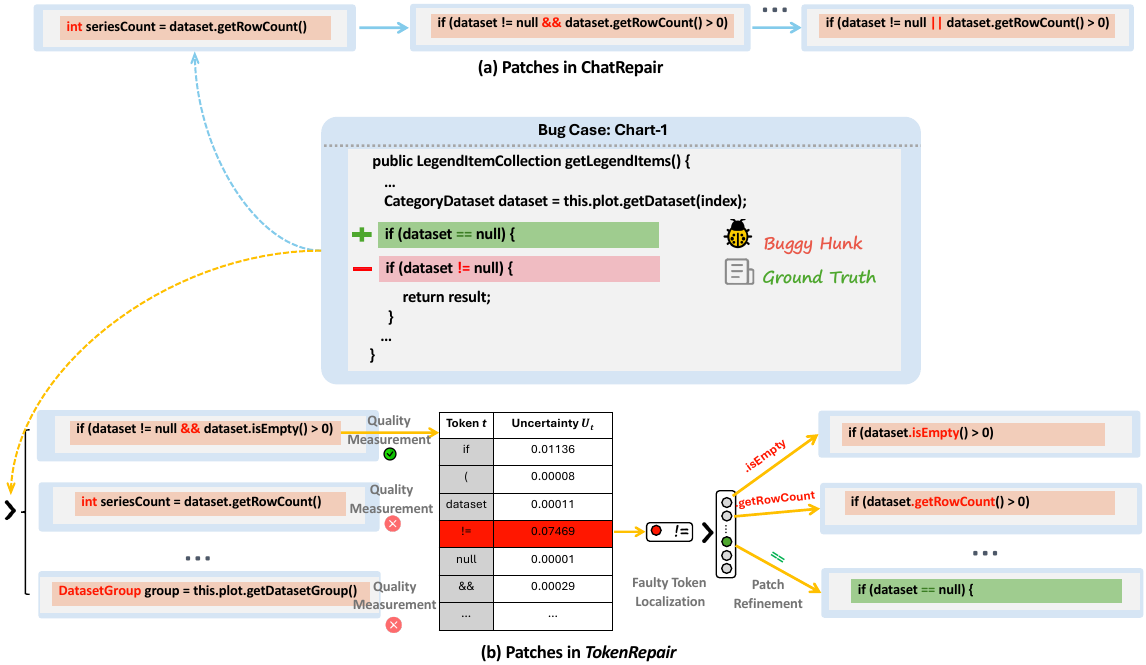}
    \vspace{-2pt}
    \caption{The motivation example}
    \label{fig:motivation}
    \vspace{-4pt}
\end{figure}
We illustrate the motivation of this work using the Chart-1 bug from the Defects4J benchmark, as shown in Fig.~\ref{fig:motivation}. The bug occurs in the \texttt{getLegendItems()} method. The original buggy condition mistakenly uses the operator ``\texttt{!=}'' to check the validity of the dataset. However, the correct logic should trigger only when the dataset is \texttt{null}, as shown in the ground-truth patch.

Recent LLM-based APR approaches attempt to fix this bug through iterative refinement~\cite{xia2023automated,10.1145/3719345}. Starting from the buggy version, the model generates a patch, tests it, and then uses the feedback to produce a new candidate. While this conversational workflow can sometimes lead to improved patches, it still struggles on cases like Chart-1. These approaches still have the following limitations:
%Recent conversation-based APR techniques typically employ an iterative repair strategy: the model generates a patch, and upon failure, treats that patch as the new buggy program, incorporates runtime or test feedback, and queries the model to produce another patch candidate. While this workflow has yielded promising results given sufficient computational resources, it exhibits two fundamental limitations that hinder repair efficacy.

\textit{1) Unaware of the faulty token.}
During patch generation, the model treats all token positions equally, without explicitly prioritizing those most likely to be faulty. Consequently, the model may modify arbitrary tokens rather than the true erroneous ones. This often produces plausible-looking but incorrect patches, causing the repair process to drift away from the actual error.
As shown in Fig.~\ref{fig:motivation}(a), starting from the buggy program, the system produces three successive patch candidates, yet all remain incorrect. None of them recognize that the real fault lies in the operator token ``\texttt{!=}''. Without explicit token-level fault localization, the model modifies irrelevant tokens, failing to replace ``\texttt{!=}'' with the correct ``\texttt{==}''. Hence, the absence of token-level guidance causes the repair process to wander rather than converge.

\textit{2) Error propagation across patches.}
Most conversational APR systems (e.g., ChatRepair) refine each generated patch unconditionally, without first assessing whether it merits further repair. This depth-first exploration strategy has two negative effects: (i) it wastes substantial computational effort refining low-quality candidates, and (ii) it allows errors in poor candidates to propagate into subsequent generations.
As shown again in Fig.~\ref{fig:motivation}(a), all three generated traces consistently preserve the expression \texttt{dataset.getRowCount()}. Because the system fails to identify these traces as low-quality, it continues building upon them, reinforcing the same incorrect logic. Once committed to an erroneous repair direction, subsequent iterations remain biased rather than exploring more promising alternatives.

Our proposed approach, \tool, addresses both limitations, as illustrated in Fig.~\ref{fig:motivation}(b). First, it performs a trace-level quality assessment to determine whether to retain each generated patch. Among all candidates, the patch \texttt{if (dataset != Null \&\& dataset.isEmpty() > 0)}
 is identified as high quality and kept for subsequent repair, while low-quality traces are filtered out. Second, \tool performs token-level fault localization within the retained patch. In particular, it identifies the operator token ``\texttt{!=}'' as the suspicious faulty token. Leveraging the model’s token-level probability distribution at that position, \tool proposes several high-probability repair candidates (e.g., replacing ``\texttt{!=}'' with ``\texttt{==}'', which ranks among the top alternatives for that position). This guided replacement is fundamentally different from unguided decoding: it explicitly targets the token most likely to cause the bug.
By combining trace-level quality assessment (to prevent error propagation) with token-level fault localization (to pinpoint and repair the true fault), \tool achieves more efficient and reliable program repair.

\section{Methodology}
\begin{figure}[t]
    \centering
    \includegraphics[width=0.9\linewidth]{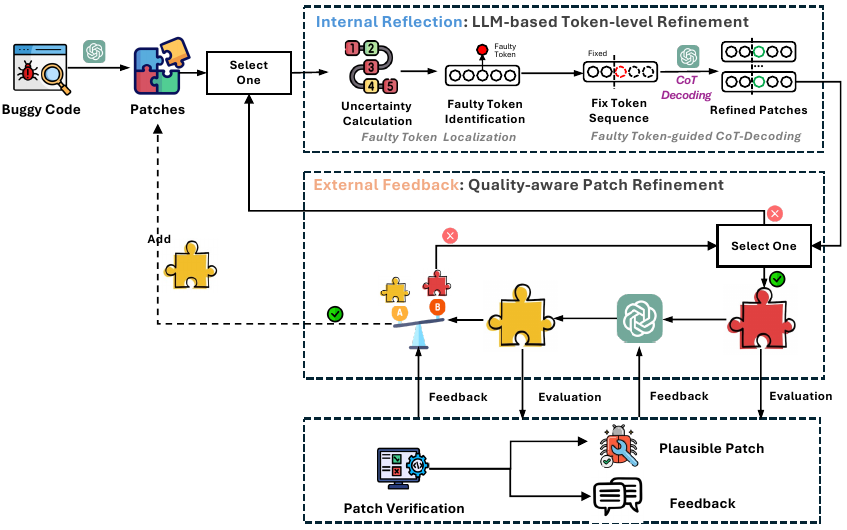}
    % \vspace{-3mm}
    \caption{Overview of \tool}
    \label{fig:overview}
    \vspace{-4pt}
\end{figure}
% In this section, we delve into the details of our proposed method \tool. We begin by providing an overview of the method and its workflow. Subsequently, we elaborate on the design principles of the two parts, internal reflection-based patch refinement and external feedback-based patch generation, consisting of four main components: faulty token localization, token-guided CoT-Decoding, feedback-based repair and trace quality measurement.

% \subsection{Overview}
Fig.~\ref{fig:overview} provides an overview of \tool. 
The approach alternates between two complementary reasoning phases: Internal reflection-based token refinement and external feedback-based patch refinement.
Given a buggy program, \tool begins by prompting an LLM to generate initial patch candidates using the buggy code and failure information (e.g., test error messages). Each generated patch then enters the internal-reflection stage, which performs token-level refinement. Specifically, \tool first conducts token-level fault localization by computing the uncertainty associated with each generated token. Suspiciousness scores are derived from the relative fluctuations in uncertainty across the token sequence, enabling \tool to identify the token positions most likely responsible for the incorrect behavior.

Once a faulty token is localized, \tool performs token-guided Chain-of-Thought (CoT) decoding. The tokens preceding the suspicious position are preserved, and new token sequences are generated from that position onward using the LLM’s probability distribution. At the localized position, \tool samples several high-probability candidate tokens and expands each into a complete patch via greedy decoding, producing a refined set of patches that specifically target the most uncertain or potentially faulty region. The key idea mirrors human reasoning: identify the uncertain steps, and explore multiple plausible alternatives to refine the solution.

If a refined patch remains incorrect after token-level internal reflection, this indicates that internal signals alone are insufficient for further improvement. In such cases, \tool transitions to the external feedback phase, where additional runtime hints (e.g., test outcomes) guide further refinement. Specifically, \tool randomly selects a candidate patch and executes it to obtain external feedback such as failure messages, stack traces, or test assertion errors. This feedback, together with the current patch, is then provided to the LLM to generate a refined patch.

The newly generated patch is subsequently evaluated along two dimensions: plausibility (i.e., whether it passes the tests) and quality (i.e., whether its uncertainty profile suggests it is a low-quality candidate). If the patch is implausible but not low-quality, it is added back into the internal-reflection pool for further token-level refinement. Conversely, if the patch is identified as low-quality, \tool discards it, since further refinement on such unstable candidates is likely to propagate errors and drift even farther from the correct fix. \tool then selects another candidate patch and repeats the external-feedback refinement process until no suitable candidates remain. Once no candidate can be selected, \tool switches back to the internal reflection phase to continue targeted token-level refinement.
This mechanism prevents the repair loop from being dominated by misleading or unreliable patches. This design mirrors human reasoning: when humans are stuck after internal deliberation, external feedback often provides new insights or constraints that help them refine their solutions more effectively. 
\subsection{Internal Reflection-based Token Refinement} \label{sec:internal}

\subsubsection{Uncertainty-guided Faulty Token Identification}\label{sec:faultytokenloc}
To identify erroneous tokens, we propose an uncertainty-guided fault localization method that estimates the model’s confidence at each token position within a generated patch. 
Specifically, for each token, we compute its uncertainty value as well as its fluctuation ratio relative to the preceding position.
Positions where uncertainty rises are marked as suspicious and considered potential fault locations.
Each suspicious token is then assigned a composite suspiciousness score that combines its uncertainty and its increase ratio, and tokens with the highest scores are selected as candidates for targeted refinement.
%Positions exhibiting uncertainty increases are flagged as suspicious and treated as potential fault locations, and each suspicious token receives a composite suspicious score derived from both its uncertainty value and the increasing ratio. Then tokens with the highest scores will be selected as candidates for targeted refinement.

%In this section, we first outline our methodology for quantifying uncertainty, and then demonstrate how it is applied to localize faulty token. To calculate token-level uncertainty, we follow the probability difference–based uncertainty method~\cite{geng2024survey} as our primary measurement approach. A detailed description of this method is provided below.\\
%\textbf{Probability Difference-based Uncertainty Computation}. 
To measure token-level uncertainty, we adopt the probability difference–based uncertainty metric~\cite{geng2024survey} as our primary indicator.
Given a prompt $x$, we utilize $U(n, x)$ to measure the uncertainty of the token at the position $n$, and the formula is defined as:  
\begin{equation}
U(n, x) = 1 - \big(\max_{y \in V}P(y \mid x, y_1, \ldots, y_{n-1}) - \max_{y \in V \backslash t^*}P(y \mid x, y_1, \ldots, y_{n-1})\big)
\label{eq:uncertainty}
\end{equation}
where $V$ is the vocabulary, $n$ represents the given position among the generated trace, and $t^*=\arg\max_{y \in V} P(y \mid x, y_1, \ldots, y_{n-1})$ is the most probable next token. Intuitively, this metric captures the gap between the top-1 and top-2 predicted token probabilities at position 
$n$. A smaller gap (i.e., larger uncertainty value) indicates that the model is less confident about the next token, signaling potential unreliability at that position.

%The larger the value of $U_d(n, x)$, the less confident the LLM is in generating the token at the position $n$, which indicates model's insufficient reasoning about the bug-related semantics and, at the same time, reflects a higher likelihood that the generated token at that position tend to be erroneous.
The value of $U(n, x)$ ranges from 0 to 1. When the model is highly confident at position $n$, its probability uncertainty is sharply peaked (e.g., $(1.0, 0.0, \ldots, 0.0)$), leading to $U(n, x) = 0$. Conversely, when uncertainty is high ($U(n, x)$ = 1), the probability distribution tends to approximate a uniform distribution, e.g., $\left( \tfrac{1}{\left| V \right|}, \tfrac{1}{\left| V \right|}, \cdots, \tfrac{1}{\left| V \right|} \right)$, where $\left| V \right|$ represents the vocabulary size of the tokenizer.

After computing the uncertainty value at each position in a trace following Equation~\ref{eq:uncertainty}, we identify all the suspicious positions by comparing each token's uncertainty with that of the preceding position. 
If the uncertainty value is larger than its predecessor's (e.g., $U(n, x) > U(n-1, x)$), the token at position $n$ is marked as suspicious. We propose a method to assign a local suspiciousness score defined as:
\begin{equation}
S_{l}(n)= U(n, x) \cdot \log \frac{U(n, x)}{U(n-1, x)} 
\label{eq:local_score}
\end{equation}
This score considers both the raw uncertainty value and the relative increase compared to the previous position, capturing both the absolute and differential aspects of uncertainty.
To further prioritize likely fault locations, we apply an exponential decay factor to each token’s local suspiciousness score based on its position in the sequence. This adjustment follows the following intuition:
% This adjustment follows the intuition that early-generation tokens are more likely to be the root cause of downstream errors, as generation mistakes often propagate forward.
when two positions have approximately equal local suspiciousness scores, we prefer to repair the earlier-positioned token first to avoid missing the truly faulty one. Thus we introduce a decay factor $\alpha$, which gradually reduces the influence of tokens appearing later in the sequence, yielding the global suspiciousness score:
%This suspicious score is then subject to exponential decay based on the token's position in the sequence, implementing the principle that earlier tokens are more likely to be the root cause of errors since generation errors tend to propagate forward. The decay factor $\alpha$ progressively reduces the suspicious scores of tokens that appear later in the sequence, ensuring that the algorithm prioritizes early suspicious tokens to be the first faulty one. 
\begin{equation}
S_{g}(n)= S_{l}(n) \cdot \alpha^n
\label{eq:global_score}
\end{equation}
Finally, the $TopK$ tokens with the highest suspiciousness scores are selected for refinement.

\subsubsection{Token-guided CoT-Decoding }
After localizing the potential faulty tokens, \tool refines them through token-guided CoT-Decoding, which preserves the content preceding the identified token within the trace and regenerates the content from that position onward.

Given the set of potential faulty tokens $FTokens = \{f_1, f_2, \ldots, f_K\}$, the refinement procedure operates directly at the token level within a candidate patch. For each faulty token $f_j$, we denote its position index in the patch's token sequence $T = (t_1, t_2, \ldots, t_L)$ as $pos(f_j)$. The prefix up to this position is preserved, i.e., $(t_1, \ldots, t_{pos(f_j)-1})$. At the faulty position $pos(f_j)$, we query the model's probability distribution $P(\cdot \mid t_{1:pos(f_j)-1})$ and identify the $m$ most probable tokens as replacement:
% \begin{equation}
% \mathcal{R}(f_j) = \{r_1, r_2, \ldots, r_m\}, \quad r_i = \argmax_{\substack{t \in V \\ t \neq f_j}} P(t \mid t_{1:pos(f_j)-1}),
% \end{equation}
% \begin{equation}
% \mathcal{R}(f_j) = \{r_1, r_2, \ldots, r_m\} = \underset{\substack{t \in V \\ t \neq f_j}}{\arg\max^{(m)}} P(t \mid t_{1:\text{pos}(f_j)-1})
% \end{equation}

% \begin{equation}
% \mathcal{R}(f_j) = \{r_1, r_2, \ldots, r_m\} = \underset{t \in V \setminus \{f_j\}}{\arg\max^{(m)}} P(t \mid t_{1:\text{pos}(f_j)-1})
% \end{equation}
\begin{equation}
\mathcal{R}(f_j) = \{r_1, r_2, \ldots, r_m\}=\underset{\substack{S \subset V \setminus \{f_j\} \\ |S| = m}}{\arg\max} \sum_{t \in S} P(t \mid t_{1:\text{pos}(f_j)-1})
\end{equation}
where $V$ is the vocabulary. For each $r_i \in \mathcal{R}(f_j)$, we construct a new patch candidate by substituting $f_j$ with $r_i$ and greedily decoding the remainder of the sequence from position $pos(f_j) + 1$. This produces $m$ refined patches per selected faulty token. Formally, the refined set for token $f_j$ is:
\begin{equation}
\mathcal{S}(f_j) = \left\{ \left(t_1, \ldots, t_{pos(f_j)-1}, r_i, \hat{t}_{pos(f_j)+1:L'}^{(i)}\right) \mid r_i \in \mathcal{R}(f_j) \right\},
\end{equation}
where $\hat{t}_{pos(f_j)+1:L'}^{(i)}$ denotes the continuation generated by greedy search conditioned on the prefix $(t_1, \ldots, t_{pos(f_j)-1}, r_i)$. The overall refined patch set is then:
\begin{equation}
\mathcal{S}_{refined} = \bigcup_{j=1}^{K} \mathcal{S}(f_j).
\end{equation}
% This refinement strategy generates a total of $K \times m$ new candidate patches, each representing a targeted correction at a specific token position.

\subsection{Fault-Prone Initial Tokens Handling}\label{sec3.2}
While the uncertainty-based method localizes erroneous tokens within a decoding sequence, it inherently relies on contextual information (i.e., previous token).
As a result, it cannot be applied to the first token, where no preceding context exists and thus no meaningful suspiciousness score can be computed.
To address this, we employ a majority voting strategy, motivated by prior findings that token frequency is positively correlated with correctness under self-consistency decoding~\cite{wang2022self}.
Specifically, we analyze the frequency distribution of candidate tokens that appear at the first position across multiple decoded traces.
We treat the occurrence frequency of each candidate token as its suspiciousness score: tokens with lower frequency are considered more likely to be incorrect. In \tool, we simply select the token with the highest occurrence frequency as the most reliable choice, while treating all remaining candidates as potentially faulty.

%However, this method has a limitation: it depends on context awareness and cannot be used to identify the correctness of the first token due to the absence of decoded context. Therefore, for distinguishing the correctness of the starting position, we adopt majority voting, following evidence that frequency correlates with correctness in self-consistency~\cite{}. Specifically, we count the frequency of possible tokens at the first position across multiple traces, treating the most frequently occurring token as correct and the remainder as incorrectly decoded tokens.

% Due to the absence of decoded context, majority voting is applied to determine the correctness of the token at the beginning of a patch. Specifically, we analyze the frequency distribution of candidate tokens at the initial position across a set of patches, identifying the token with the highest frequency as correct while categorizing all remaining tokens as decoding errors. The details of this process are as follows:
% \textbf{Majority Voting}. In standard majority voting, each trace candidate contributes equally to the final selected result. Our methods utilizes majority voting to decide the first token of a patch, because patches are worth further refinement only when the first token is correct; otherwise, the patch is temporarily waived and will be considered to be improved with low priority among the patch candidates.
Formally, let $P$ denote the set of patch candidates, and for each $p \in P$, $\texttt{FT}(p)$ denote the first token of patch $p$. The vote count for each candidate token $t$ is computed as:
\begin{equation}
V(t) = \sum_{p \in P} I(\texttt{FT}(p) = t), 
\end{equation}
where $I(\cdot)$ is the indicator function. In the token-level refinement, the final first token is determined as the one receiving the highest vote count:
\begin{equation}
\hat{t} = \arg\max_{t} V(t).
\end{equation}

After determining the first token $\hat{t}$, we proceed to apply the internal-reflection procedure (Section~\ref{sec:internal}) to identify and refine faulty tokens in the remainder of the sequence.

% all patches starting with $\hat{t}$ are considered to have the correct first token, while those with differing first tokens are identified to be erroneous and then discarded.
%Once the first token $t$ is identified, patches starting with it will be retained for further refinement, otherwise they will be waived.

\subsection{External Feedback-based Patch Refinement}
\subsubsection{Feedback-based Repair}
If internal reflection-based patch refinement still fails to generate a plausible patch, \tool attempts to leverage external feedback to further guide repair. For those refined patches which remain incorrect, they are treated as buggy code for subsequent repair iterations. The prompt is updated from $x$ to $x'$ by incorporating the refined patch and its corresponding runtime feedback, and the LLM is queried by $x'$ to explore deeper regions of the patch space, thereby augmenting the candidate pool with new patch variants.

\subsubsection{Trace Quality Measurement}\label{sec3.3.2}
Patches generated from feedback-guided repair could be of low quality, errors within them may 
propagate during further repair iterations. Therefore, trace quality measurement is performed on these newly generated patches, with only high-quality candidates being retained and added to the candidate pool.
% we introduce uncertainty-based prompt adaptation, which aims to decide whether the prompt should be updated for subsequent querying. If the new prompt could enhance the LLM's reasoning capabilities, deepen its comprehension of bug semantics while demonstrating improved trace quality, it will be retained for subsequent repairs.
Specifically, consider the refined patch and the newly generated patch derived from the prompt $x$ and $x'$, we use the uncertainty of the first token in the trace to represent its overall uncertainty. This derives from the insight: the first token of the patch carries high information content, as it establishes the syntactic and semantic context for all subsequent tokens.
Then the uncertainty values of both the two patches are calculated, which can be expressed by $U(1, x)$ and $U(1, x')$, respectively. To evaluate the quality of the new patch, we compare the two values, if $U(1, x')<U(1, x)$, the new patch will be considered as a high-quality one and retained.
% Formally, they are computed using Equation~\ref{eq:uncertainty} and can be formulated as:
% \begin{equation}
% u_{\mathcal{P}} = U_d^{\mathcal{P}}(n=1), \quad u_{\mathcal{P}'} = U_d^{\mathcal{P}'}(n=1).
% \end{equation}
% The quality classification is then determined by:
% \begin{equation}
% Q({\mathcal{P}'}) = \begin{cases}
% High & \text{if } u_{\mathcal{P}'} < u_{\mathcal{P}}, \\
% Low & \text{otherwise}.
% \end{cases}
% \end{equation}
This component is grounded in two key intuitions: 1) a patch exhibiting lower uncertainty correlates with higher quality and could increase the likelihood of successful repair; 2) a reduction in trace uncertainty relative to its predecessor indicates that the repair trajectory is progressing in the correct direction.

\subsection{Algorithm of \tool}
\begin{algorithm}
\caption{\tool}
\small
\label{algorithm:bfs}
\begin{algorithmic}[1]
\Require $b$: original buggy code, $t$: test suites, $n$: number of generated samples per query, $m$: number of generated patches by each refinement, $budget$: number of generated patches in total, $M$: LLM, $TopK$: number of selected faulty tokens, $\alpha$: decay factor
\Ensure $Patches$: the plausible patch(es)
\State $Patches := \emptyset$; $total := 0$ \label{algo2_1}
\State $f, s := \text{EvaluateToGetFeedback}(b, t)$; $prompt := \text{ConstructPrompt}(b, f)$ \label{algo2_2}
\State $P_{\text{Candis}}, U_{\text{Candis}} := \text{GeneratePatches}(M, prompt, n)$ \label{algo2_3}
\State $F, S := \text{EvaluateToGetFeedback}(P_{\text{Candis}}, t)$ \label{algo2_4}
\If{$\text{True} \in S$} \label{algo2_5}
    \State $Patches := \text{SelectValidPatches}(S, P_{\text{Candis}})$; \Return $Patches$ \label{algo2_6}
\EndIf \label{algo2_7}
\State $total := total + n$; $P_{\text{Candis}} := \text{MajorityVoting}(P_{\text{Candis}})$; $U_{\text{Candis}} := \text{GetUncert}(P_{\text{Candis}}, U_{\text{Candis}})$ \label{algo2_8}
\While{$total \leq budget$ \textbf{and} $P_{\text{Candis}} \neq \emptyset$} \label{algo2_9}
    \State $p_{\text{candi}} := \text{Pop}(P_{\text{Candis}})$; $U_{\text{candi}} := \text{Pop}(U_{\text{Candis}})$; $T := \text{GetTokens}(p_{\text{candi}})$ \label{algo2_10}
    \State $FTokens := \text{UncertaintyGuidedLocalization}(T, U_{\text{Candi}}, TopK, \alpha)$ \label{algo2_11}\BlueComment{faulty token localization}
    \State $P_{\text{Refines}}, U_{\text{Refines}} := \text{CoTDecoding}(FTokens, p_{\text{candi}}, M, m)$ \BlueComment{patch refinement} \label{algo2_12}
    \State $total := total + TopK \times m$ \label{algo2_13}
    \State $F, S := \text{EvaluateToGetFeedback}(P_{\text{Refines}}, t)$ \label{algo2_14}
    \If{$\text{True} \in S$} \label{algo2_15}
        \State $Patches := \text{SelectValidPatches}(S, P_{\text{Refines}})$; \Return $Patches$ \label{algo2_16}
    \EndIf \label{algo2_17}
    \State $P_{\text{Tmp}} := \emptyset$; $U_{\text{Tmp}} := \emptyset$ \label{algo2_18}
    \For{$p_{\text{refine}}, U_{\text{refine}}, f \in (P_{\text{Refines}}, U_{\text{Refines}}, F)$} \label{algo2_19}
        \State $prompt := \text{ConstructPrompt}(p_{\text{refine}}, f)$ \label{algo2_20}
        \State $P_{\text{Furthers}}, U_{\text{Furthers}} := \text{GeneratePatches}(M, prompt, n)$ \label{algo2_21}
        \State $F', S' := \text{EvaluateToGetFeedback}(P_{\text{Furthers}}, t)$ \label{algo2_22}
        \If{$\text{True} \in S'$} \label{algo2_23}
            \State $Patches := \text{SelectValidPatches}(S', P_{\text{Furthers}})$;
            \Return $Patches$ \label{algo2_24}
        \EndIf \label{algo2_25}
        \State $total := total + n$ \label{algo2_26}
        \For{$p_{\text{further}}, U_{\text{further}} \in (P_{\text{Furthers}}, U_{\text{Furthers}})$} \label{algo2_27}
            \State $Q := \text{MeasureTraceQuality}(p_{\text{refine}}, U_{\text{refine}}, p_{\text{further}}, U_{\text{further}})$ \BlueComment{measure trace quality} \label{algo2_28}
            \If{$Q$ is High} \BlueComment{retain high quality trace} \label{algo2_29}
                \State $P_{\text{Tmp}} := P_{\text{Tmp}} \cup p_{\text{further}}$; $U_{\text{Tmp}} := U_{\text{Tmp}} \cup U_{\text{further}}$ \label{algo2_30}
            \EndIf \label{algo2_31}
        \EndFor \label{algo2_32}
    \EndFor \label{algo2_33}
    \State $P_{\text{Tmp}} := \text{MajorityVoting}(P_{\text{Tmp}})$; $U_{\text{Tmp}} := \text{GetUncert}(P_{\text{Tmp}}, U_{\text{Tmp}})$ \label{algo2_34}
    \State $P_{\text{Candis}} := P_{\text{Candis}} \cup P_{\text{Tmp}}$; $U_{\text{Candis}} := U_{\text{Candis}} \cup U_{\text{Tmp}}$ \label{algo2_35}
\EndWhile \label{algo2_36}
\State \Return $Patches$ \label{algo2_37}
\end{algorithmic}
\end{algorithm}
% \vspace{-3pt}

The details of \tool are shown in Algorithm~\ref{algorithm:bfs}, which presents a multi-turn framework integrating both internal reflection–based refinement and external feedback–based repair within a budget-constrained loop. The framework employs a breadth-first search strategy to systematically traverse candidates,
and its detailed procedure proceeds as follows. The algorithm first initializes variables and constructs the initial prompt (Line~\ref{algo2_1} -~\ref{algo2_2}), then queries the LLM to generate $n$ patch candidates with their corresponding token-level uncertainty scores (Line~\ref{algo2_3}). Each candidate is subsequently evaluated against the test suite to obtain runtime feedback (Line~\ref{algo2_4}). If any plausible patch is identified, the algorithm terminates immediately (Line~\ref{algo2_5} -~\ref{algo2_7}).
When all initial candidates fail validation, majority voting is employed to identify promising candidates whose first tokens are 
deemed correct (Line~\ref{algo2_8}). The core iterative loop then begins (Line~\ref{algo2_9} -~\ref{algo2_36}). For each retained candidate, the algorithm first performs token-level fault localization to identify potentially erroneous tokens (Line~\ref{algo2_10} -~\ref{algo2_11}), followed by token-guided CoT-Decoding to generate refined patch variants (Line~\ref{algo2_12}). These refined patches are evaluated on the test suite (Line~\ref{algo2_14} -~\ref{algo2_17}), and the search stops immediately upon discovery of a plausible patch. 

For refined patches that remain incorrect, the algorithm conducts deeper exploration by treating each as a new buggy code instance (Line~\ref{algo2_19} -~\ref{algo2_33}). Updated prompts are constructed from the refined patch and its runtime feedback (Line~\ref{algo2_20}), followed by generation (Line~\ref{algo2_21}) and evaluation (Line~\ref{algo2_22} -~\ref{algo2_25}) of new candidates. Next, the algorithm then performs uncertainty-based quality assessment for each newly generated patch (Line~\ref{algo2_27} -~\ref{algo2_28}): patches exhibiting decreased uncertainty relative to their predecessor are classified as high quality and are temporarily retained (Line~\ref{algo2_29} -~\ref{algo2_31}). After all refined patches at the current BFS depth are explored, majority voting is applied to the temporary pool (Line~\ref{algo2_34}), and and high-quality candidates are promoted to the main candidate list for subsequent iterations (Line~\ref{algo2_35}). This iterative process continues until either a plausible patch is found or the computational budget is exhausted.

\section{Study Design}
In this section, we aim to answer the following research questions:
\begin{itemize}[leftmargin=*]
\item
\textbf{RQ1:} \textit{How do selected metrics correlate with correctness of the results?} We conduct a study to investigate the relation between metrics and patch correctness, including as the follows: 1) accuracy of faulty token localization (non-first tokens); 2) accuracy of faulty token localization (first tokens); 3) correlation between patch quality and patch correctness.

\item
\textbf{RQ2:} \textit{How effective is \tool compared with the state-of-the-art APR techniques?} We evaluate the repair performance of \tool against several existing LLM-based APR baselines to assess its overall effectiveness.

\item
\textbf{RQ3:} \textit{What are the contributions of different components of \tool in improving repair effectiveness?} We perform ablation studies to examine the impact of key modules in \tool, including majority voting-based faulty token identification, uncertainty-guided faulty token localization and trace quality measurement.
%We aim to understand the usefulness of the key components in \tool, including majority voting-based faulty token identification, uncertainty-guided faulty token localization and trace quality measurement.

\item
\textbf{RQ4:} \textit{How do different hyperparameters affect the performance of \tool?} We examine the impact of various hyperparameters on \tool's effectiveness, including the number of returned samples per query ($n$), and the number of generated patches by each refinement ($m$).
\end{itemize}

\subsection{Setup}
\subsubsection{Configuration}
We employ five pre-trained LLMs for the evaluation: Qwen2.5-Coder-7B-Instruct~\cite{hui2024qwen2, qwen2.5}, Llama-3.1-8B-Instruct~\cite{dubey2024llama, llama3.1}, DeepsSeek-Coder-6.7b-Instruct~\cite{guo2024deepseek, deepseek6.7}, DeepsSeek-Coder-7b-Instruct-V1.5~\cite{guo2024deepseek, deepseek7}, and CodeGemma-7b-it~\cite{team2024codegemma, codegemma}.
%For the experiments conducted in this paper, we opted to utilize five models, Qwen2.5-Coder-7B-Instruct~\cite{hui2024qwen2, qwen2.5}, Llama-3.1-8B-Instruct~\cite{dubey2024llama, llama3.1}, DeepsSeek-Coder-6.7b-Instruct~\cite{guo2024deepseek, deepseek6.7}, DeepsSeek-Coder-7b-Instruct-V1.5~\cite{guo2024deepseek, deepseek7}, and CodeGemma-7b-it~\cite{team2024codegemma, codegemma} as the pre-trained Large Language Model (LLM) for \tool. 
To enhance patch diversity and enlarge the search space, we followed ChatRepair to set the sampling temperature $t$ to 1. The overall patch generation budget is capped at 50, meaning that the repair process terminates once either all budgets are consumed or a plausible patch is found.
%In order to increase the potential search space and generate diverse patches, we followed ChatRepair to set the sampling temperature $t$ to 1, while setting the maximum number of generated patches ($budget$) to 50. 
The number of selected faulty tokens ($TopK$) and the decay factor ($\alpha$) are set to 3 and 0.5, respectively. 
This configuration provides a balance between precise fault localization and sufficient search depth (see \hyperref[sec:topk]{RQ1} for analysis).
%This decision strikes a balance between the efficiency and effectiveness in faulty token localization, while also preserving the opportunity to explore the deeper search space (see RQ4 for hyperparameter analysis).
For sampling parameters, we set the number of returned samples per query ($n$) to $\{2, 5\}$  and the number of generated patches by each refinement ($m$) to $\{3, 6, 9\}$.
% No explicit timeout is imposed during patch generation, and \tool stops automatically when either a plausible patch is produced or the patch-generation budget is exhausted.
%Additionally, in the process of querying LLMs to repair bugs, there is no timeout setting. Once $budgets$ are used up or a plausible patch is generated, the repair process stops.

\subsubsection{Benchmark}
We evaluate the effectiveness of \tool on two well-established benchmark datasets, including Defects4J~\cite{just2014defects4j} and HumanEval-Java~\cite{jiang2023impact}. Defects4J is a well-established collection of real-world bugs from 17 different projects, while \textit{HumanEval-Java} contains 163 buggy–fixed code pairs derived from classic programming problems.
We evaluate \tool under the single-hunk fix scenario, and adopt the benchmark construction process in prior research~\cite{xia2024automated, kong2025demystifying}, where the location of the buggy hunk is provided based on the ground truth.
  
For the Defects4J dataset, we target the 154 single-hunk bugs in Defects4J 1.2, and for HumanEval-Java, we assess the effectiveness of \tool on all the 163 bugs.

\subsubsection{Baselines}
%In our comparative evaluation, we evaluate \tool by comparing it with three state-of-the-art baselines, all of them are LLM-based repair tools,
We compare \tool against three representative state-of-the-art LLM-based APR baselines:
\begin{itemize}[leftmargin=*]
    \item \textbf{Base Sampling~\cite{kong2025demystifying}}. This method leverages the inherent stochasticity in outputs of the LLM by scaling up the number of samples (i.e., generating more candidates at inference time) rather than relying on external feedback, to search for optimal results. We follow the experimental setup in the original paper. During each query, the initial prompt remains fixed (i.e., no feedback-based prompt updates), while up to the full patch budget is used for sampling.
    \item \textbf{CoT-Decoding~\cite{wang2024chain}}. The method explores multiple alternative decoding paths rather than blindly following the token with the highest probability. At the first decoding step, the model looks at the $TopN$ tokens with highest probability from the vocabulary. Once the first token is identified, the trace continues with standard greedy decoding for all subsequent tokens. In the experiments, we set the number of $TopN$ paths as the same as $budget$.
    \item \textbf{ChatRepair~\cite{xia2024automated}}. 
    ChatRepair iteratively repairs code by incorporating runtime feedback in conversational turns.
    We configure this baseline to match \tool in temperature ($t$), samples per query ($n$), and patch-generation budget.
    The key distinction is that ChatRepair lacks internal reflection and relies solely on coarse-grained test outcomes as external feedback. Consequently, it directly reuses invalid patches in subsequent iterations without token-level refinement or trace-quality evaluation.
    %The baseline is configured identically to \tool with respect to temperature $t$, samples per query $n$, and maximum number of patches $budget$, and likewise employs runtime feedback for guiding iterative repair. The fundamental difference between ChatRepair and our method is that the former lacks internal reflection and it directly utilizes invalid patch candidates for subsequent repair without conducting intermediate patch refinement or trace quality measurement.
    
\end{itemize}
\subsubsection{Metrics} We select three widely-used metrics to compare \tool with the baselines:
\begin{itemize}[leftmargin=*]
    \item \textit{Number of Plausible Fixes ($\#Plausible$)}: Evaluates the capability of the repair method to generate plausible patches. It counts the number of patches that can pass all the test suites after repairing.
    \item \textit{Number of Correct Fixes ($\#Correct$)}: Assesses the ability of the repair method to produce accurate patches. This metric counts the number of programs that have been properly repaired based on a manual review of the plausible patches generated by each tool.
    \item \textit{Number of Generated Patches ($\#Generate$)}: Quantifies the resource utilization of the LLM-based method. We evaluate the average number of patch candidates generated across all bug cases which could be fixed correctly, providing insight into the efficiency of individual methods.
\end{itemize}
\section{Evaluation}
\subsection{RQ1: Correlation between Metrics and Correctness}\label{rq1}
The foundation of \tool lies in its patch refinement mechanism, which combines internal reflection and external feedback. In particular, our design introduces three key metrics that guide refinement:
(1) \textit{token-level suspiciousness scores} for localizing faulty non-first tokens,
(2) \textit{first-token majority voting} for identifying faulty initial tokens, and
(3) \textit{trace quality measurement} based on first-token uncertainty to filter low-quality patches.
These metrics play a critical role in enabling fine-grained and effective repair.

In this RQ, we aim to investigate the correlation between these metrics and patch correctness. Specifically, we evaluate: whether the token localization mechanisms can accurately identify faulty tokens in incorrect patches (including both first tokens and non-first tokens), and whether the trace quality metric is strongly correlated with patch correctness.

\phantomsection\label{sec:topk}
\textbf{Accuracy of Faulty Token Localization (Non-First Tokens).}
As described in Section~\ref{sec:faultytokenloc}, we use suspiciousness scores to localize the token that is most likely to be faulty. To evaluate the effectiveness of this mechanism, we examine all generated patches (both correct and incorrect) and check whether the token identified by our method corresponds to an actual faulty token. If the localized token is indeed faulty, we mark the case as correct localization; otherwise, it is labeled as incorrect localization. This allows us to assess how accurately the suspiciousness-based localization identifies true fault positions.

The suspiciousness score depends on two key hyperparameters: the decay factor $\alpha$ and the number of selected suspicious tokens $TopK$. To understand their impact, we conduct a grid search over $\alpha \in \{0.2, 0.5, 0.8\}$ and $TopK \in \{1, 2, 3, 4, 5\}$, evaluating localization accuracy across all five models on both benchmarks. For each hyperparameter configuration, we measure the localization performance achieved by \tool to identify the setting that most reliably detects faulty tokens.

% \textbf{Hyperparameters for Faulty Token Localization.} 
% As demonstrated in Section~\ref{sec:internal}, the decay factor $\alpha$ and the number of selected suspicious tokens $TopK$ are two critical hyperparameters in the faulty token localization component. To investigate the effectiveness of configuring them, we conducted a grid search by varying $\alpha \in \{0.2, 0.5, 0.8\}$ and $TopK \in \{1, 2, 3, 4, 5\}$, evaluating localization accuracy across all five models on both benchmarks, and we evaluated the localization performance among all the traces generated by \tool in RQ2.
\begin{table*}[t]
\centering
\caption{Localization performance across different decay factors and $TopK$ selection}
\label{tab:decay_factor_performance}
\small
\scalebox{0.85}{
\begin{tabular}{lcccccccc}
\toprule
\textbf{\textit{Defects4J}}& \textbf{Decay Factor} & \textbf{Top-1 Acc} & \textbf{Top-2 Acc} & \textbf{Top-3 Acc} & \textbf{Top-4 Acc} & \textbf{Top-5 Acc} & \textbf{Avg.} \\
\midrule
% \multicolumn{8}{l}{\textbf{Defects4J}} \\
% \midrule
& $\alpha = 0.2$ & 0.484 & 0.603 & 0.742 & 0.759 & 0.775 & 0.673 \\
Qwen& $\alpha = 0.5$ & 0.484 & 0.625 & 0.724 & 0.758 & 0.806 & \textbf{0.679} \\
& $\alpha = 0.8$ & 0.482 & 0.604 & 0.703 & 0.763 & 0.782 & 0.667 \\
\cdashline{2-8}
& $\alpha = 0.2$ & 0.449 & 0.635 & 0.752 & 0.799 & 0.838 & \textbf{0.695} \\
Llama& $\alpha = 0.5$ & 0.448 & 0.586 & 0.692 & 0.787 & 0.841 & 0.671 \\
 & $\alpha = 0.8$ & 0.447 & 0.577 & 0.662 & 0.759 & 0.803 & 0.650 \\
 \cdashline{2-8}
& $\alpha = 0.2$ & 0.448 & 0.591 & 0.717 & 0.802 & 0.831 & 0.678 \\
DeepSeek& $\alpha = 0.5$ & 0.448 & 0.659 & 0.739 & 0.793 & 0.834 & \textbf{0.695} \\
 & $\alpha = 0.8$ & 0.446 & 0.615 & 0.722 & 0.805 & 0.839 & 0.685 \\
 \cdashline{2-8}
& $\alpha = 0.2$ & 0.405 & 0.592 & 0.733 & 0.786 & 0.821 & 0.667 \\
CodeGemma& $\alpha = 0.5$ & 0.405 & 0.601 & 0.734 & 0.812 & 0.851 & \textbf{0.681} \\
 & $\alpha = 0.8$ & 0.405 & 0.559 & 0.656 & 0.746 & 0.825 & 0.638 \\
 \cdashline{2-8}
& $\alpha = 0.2$ & 0.435 & 0.592 & 0.653 & 0.713 & 0.795 & 0.638 \\
DeepSeek-V1.5& $\alpha = 0.5$ & 0.435 & 0.592 & 0.674 & 0.743 & 0.794 & \textbf{0.648} \\
 & $\alpha = 0.8$ & 0.434 & 0.569 & 0.688 & 0.747 & 0.789 & 0.645 \\
\midrule
\textbf{\textit{HumanEval-Java}}& \textbf{Decay Factor} & \textbf{Top-1 Acc} & \textbf{Top-2 Acc} & \textbf{Top-3 Acc} & \textbf{Top-4 Acc} & \textbf{Top-5 Acc} & \textbf{Avg.} \\
\midrule
& $\alpha = 0.2$ & 0.297 & 0.545 & 0.648 & 0.706 & 0.734 & 0.586 \\
Qwen& $\alpha = 0.5$ & 0.297 & 0.521 & 0.701 & 0.784 & 0.804 & \textbf{0.621} \\
 & $\alpha = 0.8$ & 0.297 & 0.536 & 0.663 & 0.748 & 0.778 & 0.604 \\
 \cdashline{2-8}
& $\alpha = 0.2$ & 0.273 & 0.463 & 0.564 & 0.629 & 0.718 & 0.529 \\
Llama& $\alpha = 0.5$ & 0.269 & 0.509 & 0.624 & 0.704 & 0.768 & \textbf{0.592} \\
 & $\alpha = 0.8$ & 0.269 & 0.474 & 0.579 & 0.719 & 0.771 & 0.562 \\
 \cdashline{2-8}
& $\alpha = 0.2$ & 0.231 & 0.441 & 0.564 & 0.660 & 0.756 & 0.530 \\
DeepSeek& $\alpha = 0.5$ & 0.231 & 0.517 & 0.639 & 0.751 & 0.804 & \textbf{0.589} \\
 & $\alpha = 0.8$ & 0.231 & 0.488 & 0.642 & 0.766 & 0.810 & 0.587 \\
 \cdashline{2-8}
& $\alpha = 0.2$ & 0.326 & 0.595 & 0.697 & 0.751 & 0.794 & \textbf{0.633} \\
CodeGemma& $\alpha = 0.5$ & 0.326 & 0.604 & 0.672 & 0.728 & 0.791 & 0.624 \\
 & $\alpha = 0.8$ & 0.325 & 0.486 & 0.606 & 0.696 & 0.759 & 0.574 \\
 \cdashline{2-8}
& $\alpha = 0.2$ & 0.337 & 0.495 & 0.584 & 0.679 & 0.755 & 0.570 \\
DeepSeek-V1.5& $\alpha = 0.5$ & 0.337 & 0.616 & 0.695 & 0.724 & 0.757 & 0.626 \\
 & $\alpha = 0.8$ & 0.337 & 0.588 & 0.686 & 0.756 & 0.803 & \textbf{0.634} \\
\bottomrule
\end{tabular}
}
\vspace{-3mm}
\end{table*}
% \vspace{-5pt}

Table~\ref{tab:decay_factor_performance} presents the $TopK$ accuracy results across different hyperparameter configurations. The results demonstrate that the suspiciousness score can be effectively leveraged to localize faulty tokens (excluding the first token), achieving average localization accuracy ranging from 0.589 to 0.695, indicating promising effectiveness of uncertainty-guided token-level fault localization, and there is a strong correlation between suspiciousness scores and patch correctness.

For the decay factor $\alpha$, we observe that intermediate values usually outperform extreme settings. When $\alpha$ is too small (0.2), the decay effect becomes overly aggressive, heavily penalizing tokens appearing later in the sequence. This can lead to underestimation of their suspiciousness and reduced accuracy, particularly noticeable in larger $K$ values, e.g., Top-5. Conversely, when $\alpha$ is too large (0.8), the decay effect diminishes, allowing later tokens to retain relatively high suspicious scores. This dilutes the prioritization of early faulty tokens and degrades localization precision, especially at smaller $K$ values, e.g., Top-1. The intermediate setting $\alpha = 0.5$ achieves optimal performance across most configurations, as reflected by the average accuracy (Avg.). For instance, on Defects4J, $\alpha = 0.5$ yields the highest average accuracy for four out of five models.
Regarding the selection of $TopK$, we observe substantial accuracy gains when increasing $K$ from 1 to 3, but diminishing returns beyond $K = 3$. For example, on Defects4J with Qwen and $\alpha = 0.5$, Top-1 accuracy is 0.484, which increases significantly to 0.625 at Top-2 and 0.724 at Top-3. However, further increases to Top-4 (0.758) and Top-5 (0.806) yield progressively weaker improvements. Similar patterns emerge across the other models and benchmark, indicating that the most of actual faulty tokens fall within the three most suspicious positions. More importantly, selecting larger $K$ values incurs substantial computational costs: the total number of generated patches will increase by $TopK$ × $m$ for refining each invalid patch candidate, where $m$ is the number of refined patch variants per suspicious token. Increasing $K$ directly reduces the remaining budget available for external feedback-based repair, which is crucial for handling complex bugs requiring multi-turn repair. Therefore, we adopt $\alpha = 0.5$ and $TopK=3$ as the default configuration for \tool.

\textbf{Accuracy of Faulty Token Localization (First Tokens).}
As described in Section~\ref{sec3.2}, majority voting serves as a critical filtering mechanism in \tool, enabling the system to identify patches with correct first tokens before proceeding to more computationally expensive refinement steps. Therefore, establishing a strong correlation between majority-vote predictions and actual first-token correctness is essential to validate this design choice.

To evaluate this correlation, we treat majority voting as a binary predictor of first-token correctness. For each patch, the majority-vote result is used to predict whether its first token is correct. The prediction is then compared against the ground truth: if the predicted correctness matches the actual correctness, the case is labeled as correct; otherwise, it is labeled as incorrect. This allows us to quantitatively assess how reliably majority voting identifies correct first tokens.

\begin{table}[t]
\centering
\caption{Correlation between majority-voting and correctness of initial token}
\label{tab:majority_voting}
\scalebox{0.9}{
\begin{tabular}{lcccccc}
\toprule
\textbf{Benchmark}& \multicolumn{3}{c}{\textbf{\textit{Defects4J}}} & \multicolumn{3}{c}{\textbf{\textit{HumanEval-Java}}} \\
\cmidrule(lr){2-4} \cmidrule(lr){5-7}
\textbf{Model} & \textbf{Precision} & \textbf{Recall} & \textbf{F1 Score} & \textbf{Precision} & \textbf{Recall} & \textbf{F1 Score} \\
\midrule
Qwen         & 0.678 & 0.912  & 0.778        & 0.863 & 0.978 & 0.917 \\
Llama        & 0.705 &	0.947 &	0.809& 	
0.843 &	0.975 &	0.904 \\
DeepSeek     & 0.507&	0.838&	0.624&	0.818&	0.964&	0.885 \\
CodeGemma    & 0.598&	0.871&	0.709&	0.788&	0.945&	0.859 \\
DeepSeek-V1.5 & 0.606&	0.892&	0.721&	0.872&	0.991&	0.928 \\
\bottomrule
\end{tabular}
}
\vspace{-3mm}
\end{table}

Table~\ref{tab:majority_voting} presents the classification performance of majority voting in terms of precision, recall, and F1 score across both benchmarks and all five evaluated models. Overall, the results demonstrate a strong positive correlation between majority voting and first-token correctness across both datasets. The recall values are consistently high, ranging from 0.838 to 0.991, indicating that majority voting successfully identifies the vast majority of patches with correct first tokens. The F1 scores, which provide a balanced assessment, range from 0.624 to 0.928, confirming that majority voting constitutes a reliable indicator for identifying the initial token. Examining individual model performance, DeepSeek-Coder-6.7b-Instruct on Defects4J presents a notable case of weaker correlation, achieving a precision of only 0.507 and an F1 score of 0.624, the lowest among all configurations. The low precision (0.507) indicates a high false-positive rate, meaning that majority voting frequently predicts a first token as correct when it is actually incorrect. This suggests that for a substantial proportion of bugs in Defects4J, DeepSeek-Coder-6.7b-Instruct generates multiple patches that consistently start with the same erroneous token, thereby misleading the majority voting mechanism. This phenomenon reflects a model-specific generation bias where certain incorrect token choices are systematically preferred due to the model's training distribution or architectural characteristics. Despite this limitation in precision, the model maintains a relatively high recall (0.838), ensuring that most truly correct first tokens are still identified. Nevertheless, the overall performance validates the feasibility of majority voting for faulty token identification. 
% Therefore, the  evidence confirms that majority voting constitutes a valid approach for first-token identification.

\textbf{Correlation between Patch Quality and Patch Correctness}.
As described in Section~\ref{sec3.3.2}, \tool leverages uncertainty trends to measure trace quality, which reflects whether the repair process is progressing toward or diverging from the correct solution. 
We investigate whether decreased uncertainty during iterative repair correlates with correct fixes. Specifically, we collected all the patches generated by \tool, and organized them into individual paths, where each path consists of a sequence of consecutive patches representing a continuous repair trajectory. These paths can be categorized into two groups: plausible paths, where a plausible patch eventually emerges, and incorrect paths, where no plausible patch is found. For each patch in a path, we compared its uncertainty with that of its predecessor, classifying the transition as either increasing ($Uncert.\uparrow$) or decreasing ($Uncert.\downarrow$). We then computed the proportion of each trend type within plausible and incorrect paths.

Table~\ref{tab:uncert_tendency_correctness} presents the proportions of decreasing and increasing uncertainty transitions across both benchmarks. For incorrect paths, we observe that the proportions of decreasing and increasing trends are relatively balanced across both benchmarks. Taking paths in Defects4J as an example, incorrect paths show nearly equal distributions: decreasing ratios range from 40.6\% to 51.9\%, while increasing ratios range from 48.1\% to 59.4\%. This near-equilibrium indicates that unsuccessful repair trajectories lack clear directional trends, with increasing and decreasing ratios showing no obvious differences. In contrast, plausible paths demonstrate a pronounced bias toward decreasing uncertainty trend. In Defects4J, decreasing tendency account for 55.8\% to 64.5\% of transitions in plausible paths, substantially exceeding the 35.5\% to 44.2\% for increasing tendencies. This pattern becomes even more striking on HumanEval-Java, where decreasing uncertainty dominates with proportions ranging from 54.1\% to 80.5\%. This consistent disparity across all models and both benchmarks indicates that successful repair trajectories are characterized by progressively increasing model confidence, as reflected by declining uncertainty values across consecutive iterations. Thus uncertainty trend constitutes a significantly effective indicator for patch quality assessment, which validates \tool's design choice to incorporate uncertainty tendency into trace quality measurement.

\begin{table*}[t]
\centering
\caption{Relation between patch quality and patch correctness}
\label{tab:uncert_tendency_correctness}
\resizebox{\textwidth}{!}{
\begin{tabular}{lcccclcccc}
\toprule
\textbf{\textit{Defects4J}}& \multicolumn{2}{c}{\textbf{Plausible Path}} & \multicolumn{2}{c}{\textbf{Incorrect Path}} & \textbf{\textit{HumanEval-Java}} & \multicolumn{2}{c}{\textbf{Plausible Path}} & \multicolumn{2}{c}{\textbf{Incorrect Path}} \\
\cmidrule(lr){2-3} \cmidrule(lr){4-5} \cmidrule(lr){7-8} \cmidrule(lr){9-10}
\textbf{Models} & \textbf{$Uncert.\downarrow$} & \textbf{$Uncert.\uparrow$} & \textbf{$Uncert.\downarrow$} & \textbf{$Uncert.\uparrow$} & \textbf{Models} & \textbf{$Uncert.\downarrow$} & \textbf{$Uncert.\uparrow$} & \textbf{$Uncert.\downarrow$} & \textbf{$Uncert.\uparrow$} \\
\midrule
Qwen& 64.5\% &35.5\%& 50.4\%& 49.6\%& Qwen& 80.5\%& 19.5\%& 60.4\%& 39.6\% \\
Llama& 58.1\%& 41.9\%& 51.9\%&	48.1\%& Llama& 77.4\%&	22.6\%& 58.4\%& 41.6\% \\
DeepSeek& 58.8\%& 41.2\%& 40.6\%& 59.4\%& DeepSeek& 54.1\%& 45.9\%& 49.9\%& 50.1\% \\
CodeGemma& 58.3\%& 41.7\%& 45.1\%&	54.9\%& CodeGemma& 65.5\%& 34.5\%& 47.8\%& 52.2\% \\
DeepSeek-V1.5& 55.8\%& 44.2\%& 45.8\%&	54.2\%& DeepSeek-V1.5& 59.1\%& 40.9\%& 40.3\%& 59.7\% \\
\bottomrule
\end{tabular}
}
\vspace{-3mm}
\end{table*}

\begin{tcolorbox}[size=title,opacityfill=0.1]
\noindent \textbf{Answer to RQ1: }
% Our investigation reveals that both trace uncertainty and majority voting correlate with the correctness in distinct ways. Specifically, uncertainty trend demonstrates that a repair trajectory which could yield a plausible patch consistently exhibit higher proportions (from 55.8\% to 80.5\%) in decreasing trend of uncertainty. 
% And majority voting could effectively identify correct initial tokens with high recall (from 0.838 to 0.991) and F1 scores (from 0.624 to 0.928).
Our results show that all three metrics exhibit strong correlations with patch correctness, which is critical in improving program repair performance. 
% For faulty token localization (excluding the first token), the suspiciousness score achieves strong accuracy, with average values ranging from 0.589 to 0.695, indicating promising effectiveness. For fault localization at the initial token position, majority voting demonstrates strong effectiveness, achieving recall values ranging from 0.838 to 0.991 and F1 scores ranging from 0.507 to 0.624. Regarding patch quality and correctness, decreasing uncertainty dominates in plausible paths (those culminating in correct patches), accounting for 54.1\% to 80.5\% of transitions—substantially higher than the proportion observed in incorrect paths.
\end{tcolorbox}

\subsection{RQ2: Effectiveness of \tool}
\begin{table*}[t]
\centering
\caption{Comparative results on Defects4J and HumanEval-Java}
\label{tab:repair_comparison}
\small
\setlength{\tabcolsep}{3.5pt}
\scalebox{0.9}{
\begin{tabular}{lcccccccccccc}
\toprule
\textbf{\textit{Defects4J}}& \multicolumn{3}{c}{\textbf{Base Sampling}} & \multicolumn{3}{c}{\textbf{CoT-Decoding}} & \multicolumn{3}{c}{\textbf{ChatRepair}} & \multicolumn{3}{c}{\textbf{\tool}} \\
\cmidrule(lr){2-4} \cmidrule(lr){5-7} \cmidrule(lr){8-10} \cmidrule(lr){11-13}
 \textbf{Models} & $\#Plaus.$ & $\# Cor.$ & $\# Gen.$ & $\#Plaus.$ & $\# Cor.$ & $\# Gen.$ & $\#Plaus.$ & $\# Cor.$ & $\# Gen.$ & $\#Plaus.$ & $\# Cor.$ & $\# Gen.$ \\
\midrule
Qwen & 67 & 51 & 13.12 & 59 & 42 & 9.03 & 61 & 51 & 9.41 & \textbf{80} & \textbf{63} & 7.50 \\
Llama & 66 & 41 & 9.22 & 35 & 25 & 10.57 & 64 & 49 & 8.33 & \textbf{72} & \textbf{53} & 9.75 \\
DeepSeek & 68 & 46 & 10.75 & 50 & 38 & 9.62 & 65 & 45 & 9.93 & \textbf{73} & \textbf{53} & 9.82 \\
CodeGemma & 60 & 37 & 11.54 & 30 & 24 & 11.03 & 59 & 43 & 9.47 & \textbf{73} & \textbf{58} & 11.08 \\
DeepSeek-V1.5 & 61 & 43 & 11.79 & 51 & 38 & 11.75 & 60 & 39 & 8.66 & \textbf{72} & \textbf{54} & 11.24 \\
\midrule
\textbf{\textit{HumanEval-Eval}} & \multicolumn{3}{c}{\textbf{Base Sampling}} & \multicolumn{3}{c}{\textbf{CoT-Decoding}} & \multicolumn{3}{c}{\textbf{ChatRepair}} & \multicolumn{3}{c}{\textbf{\tool}} \\
\cmidrule(lr){2-4} \cmidrule(lr){5-7} \cmidrule(lr){8-10} \cmidrule(lr){11-13}
\textbf{Models} & $\#Plaus.$ & $\# Cor.$ & $\# Gen.$ & $\#Plaus.$ & $\# Cor.$ & $\# Gen.$ & $\#Plaus.$ & $\# Cor.$ & $\# Gen.$ & $\#Plaus.$ & $\# Cor.$ & $\# Gen.$ \\
\midrule
Qwen & 113 & 103 & 6.87 & 108 & 95 & 6.41 & 114 & 101 & 5.71 & \textbf{124} & \textbf{110} & 5.42 \\
Llama & 108 & 92 & 8.81 & 74 & 69 & 4.64 & 107 & 90 & 7.06 & \textbf{110} & \textbf{95} & 7.62 \\
DeepSeek & \textbf{114} & \textbf{99} & 7.10 & 110 & 93 & 6.71 & 114 & 98 & 5.85 & 114 & 98 & 5.53 \\
CodeGemma & 108 & 93 & 7.82 & 83 & 72 & 5.73 & 99 & 90 & 5.43 & \textbf{114} & \textbf{99} & 5.79 \\
DeepSeek-V1.5 & 100 & 88 & 8.22 & 97 & 86 & 7.39 & 101 & 93 & 5.19 & \textbf{118} & \textbf{108} & 5.44 \\
\bottomrule
\end{tabular}
}
\end{table*}

Table~\ref{tab:repair_comparison} summarizes the comparative repair results across all models and benchmarks. Overall, \tool consistently outperforms existing approaches in most settings, achieving the highest number of correct and plausible fixes. 

On Defects4J, \tool exhibits a substantial advantage. Using Qwen as the model, it achieves 63 correct fixes, an improvement of 23.5\% over the second-best baseline, ChatRepair, which attains 51. 
Similarly, on HumanEval-Java, \tool achieves 108 correct fixes with DeepSeek-V1.5, surpassing ChatRepair’s 93 fixes by 16.1\%. 
These results demonstrate the effectiveness of \tool on program repair compared with state-of-the-art approaches.
%These results demonstrate that the combination of faulty token localization, targeted refinement, and trace quality measurement enables \tool to explore the patch space more effectively than existing approaches. 
One exception occurs on HumanEval-Java with DeepSeek, where Base Sampling slightly surpasses \tool (99 vs. 98 correct fixes). 
This marginal drop can be attributed to DeepSeek’s weaker token localization accuracy. As reported in Table~\ref{tab:decay_factor_performance}, its average localization accuracy (0.589) is the lowest among all configurations. Consequently, \tool occasionally misidentifies faulty tokens, leading to misguided refinements at irrelevant positions and inefficient use of computational budget.
%This result can be attributed to the poor faulty token localization performance of DeepSeek on HumanEval-Java. As shown in Table~\ref{tab:decay_factor_performance}, the average localization accuracy of DeepSeek is 0.589, which is the lowest value among all the configurations. This weak localization capability leads to incorrect identification of faulty tokens, causing \tool to remediate irrelevant positions and waste computational resources, ultimately resulting in slightly worse performance than Base Sampling approach. This case underscores the importance of accurate token-level fault localization as a prerequisite for effective patch refinement.

\begin{figure}[h]
    \centering
    \includegraphics[width=0.8\linewidth]{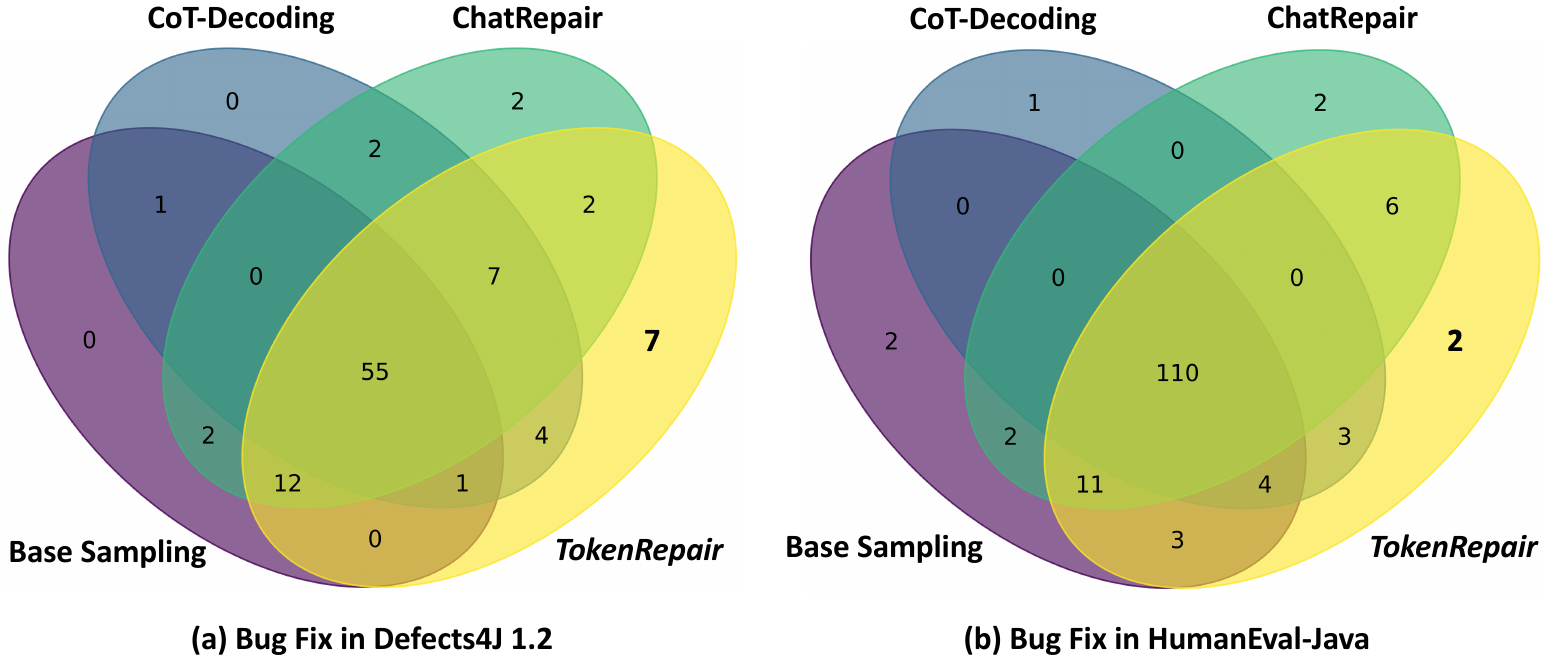}
    \caption{Bug fix Venn diagram in two benchmarks}
    \label{fig:venn}
\end{figure}

Fig.~\ref{fig:venn} presents Venn diagrams illustrating the cumulative repair capabilities when aggregating results across all five models. On Defects4J (Fig.~\ref{fig:venn}(a)), Base Sampling, CoT-Decoding, ChatRepair, and \tool collectively achieve 71, 70, 82, and 88 correct fixes across all models, respectively. The results demonstrate \tool achieves the best performance and it contributes to an improvement of 7.3\% over the second-best method ChatRepair (88 vs. 82). In addition, \tool can fix 7 unique bugs that have not been solved by the others. On HumanEval-Java (Fig.~\ref{fig:venn}(b)), the four methods achieve 132, 118, 131, and 139 correct fixes, respectively. \tool uniquely fixes 2 bugs and increases the total correct fixes by 6.1\% compared to ChatRepair (139 vs. 131). The consistent superiority across aggregated results confirms that the effectiveness of \tool generalizes well across diverse models and benchmarks.

\begin{figure}[h]
    \centering
    \includegraphics[width=0.9\linewidth]{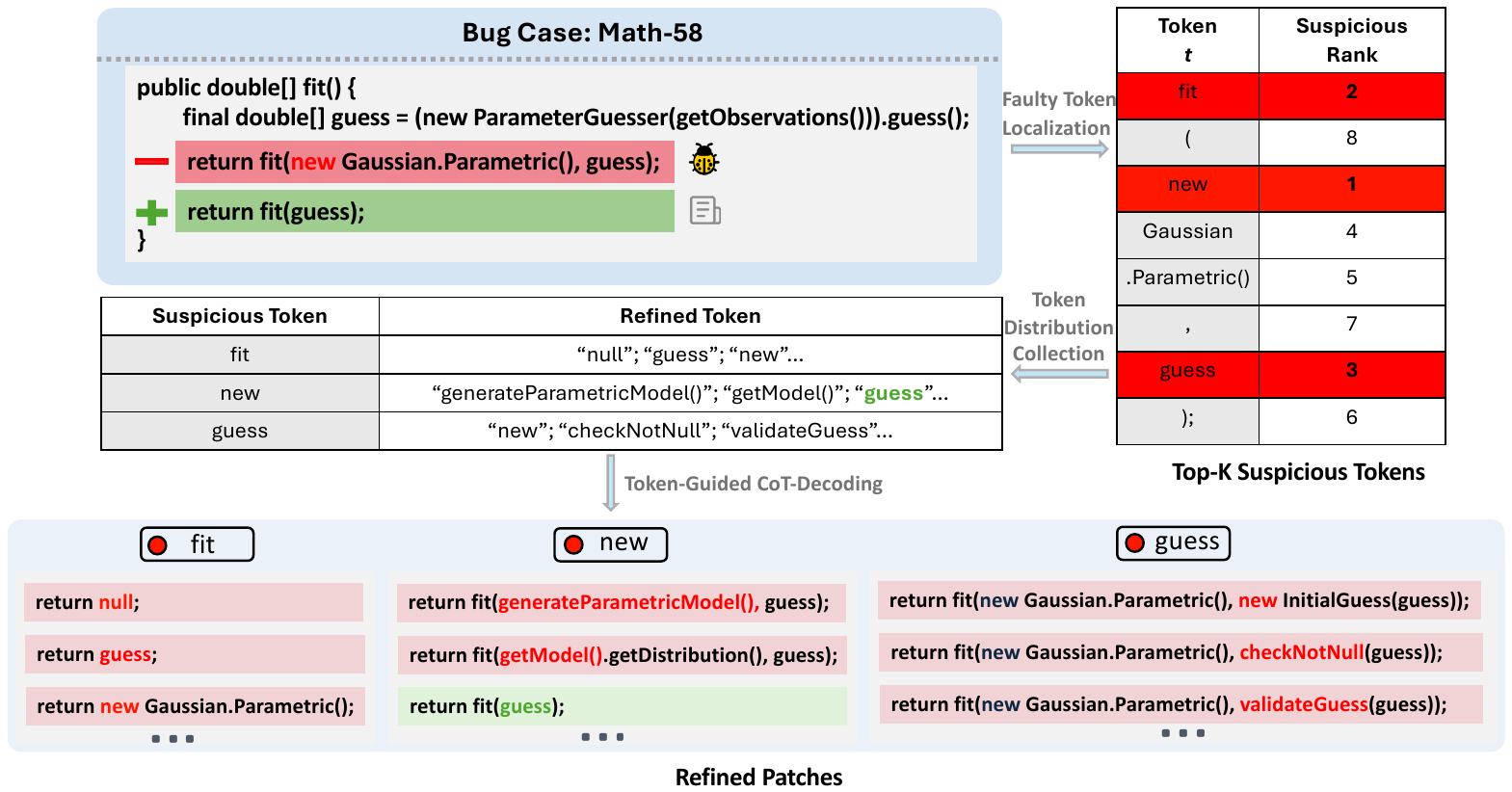}
    \caption{An illustrative example of a bug uniquely fixed by \tool in Defects4J 1.2}
    \label{fig:case}
\end{figure}

\textbf{Efficiency.}
We evaluate repair efficiency using the metric $\#Gen.$ in Table~\ref{tab:repair_comparison}, which measures the average number of patches generated per correct fix. Lower values indicate higher efficiency and reduced computational cost. For example, Qwen with \tool demonstrates the best efficiency across both benchmarks among all the methods, requiring only 7.50 patches per fix on Defects4J and 5.42 on HumanEval-Java. Compared with the most efficient baselines, the computation cost can decrease by 16.9\% (7.50 vs. 9.03) and 5.1\% (5.42 vs. 5.71), respectively. For most other settings, \tool achieves competitive efficiency with minimal cost increases compared to the most efficient baseline. For instance, on Defects4J with DeepSeek, \tool requires 9.82 patches per fix compared to 9.62 by CoT-Decoding, representing only 2.1\% increase. Similarly, DeepSeek-V1.5 shows comparable efficiency on HumanEval-Java relative to the most efficient baselines. In some cases, \tool incurs higher computational costs, such as Llama on HumanEval-Java where the cost increases from 4.64 to 7.62 compared to CoT-Decoding. However, this efficiency trade-off is justified by substantial effectiveness gains: Llama fixes 95 bugs with \tool versus 69 with CoT-Decoding, a 37.7\% improvement. The increased cost reflects the ability to successfully repair challenging bugs requiring extensive costs near the budget limit, which the other methods cannot resolve. Overall, \tool maintains competitive efficiency while delivering superior effectiveness.

\textbf{Case Study.} 
Fig.~\ref{fig:case} presents the detailed repair process for bug case \texttt{Math-58} from Defects4J, which is an unique fix by \tool. The original buggy code incorrectly returns \texttt{fit(new Gaussian.Parametric(), guess)}, while the ground truth requires simply returning \texttt{fit(guess)}. After collecting uncertainty values to compute suspiciousness scores at each token position, all positions are ranked by the scores. \tool identifies the most suspicious tokens: ``\texttt{new}'' (rank 1), ``\texttt{fit}'' (rank 2), and ``\texttt{guess}'' (rank 3). According to the hyperparameter configuration  $TopK = 3$, they are selected as potential faulty tokens for subsequent refinement. For each identified faulty token position, \tool queries the model's output probability distribution over the vocabulary and extracts $m$ candidate tokens with the highest-probability for replacement. Through token-guided CoT-Decoding, each candidate token serves as the starting point for greedy generation of the remaining sequence, producing refined patch variants. For instance, at the position of ``\texttt{new}'', the three candidate tokens include``\texttt{generateParametricModel()}'', ``\texttt{getModel()}'', and ``\texttt{guess}'', leading to refined patches: \texttt{return fit(generateParametricModel(), guess)}, \texttt{return fit(getModel().getDistribution(), guess)}, and \texttt{return fit(guess)}, respectively. Among all the refined patches generated through this systematic exploration, \texttt{return fit(guess)} emerges as the correct solution.

\begin{tcolorbox}[size=title,opacityfill=0.1]
\noindent \textbf{Answer to RQ2: }
\tool demonstrates superior repair effectiveness across both benchmarks. 
% It totally achieves 88 correct fixes on Defects4J and 139 on HumanEval-Java. 
Compared to the best-performing baseline, \tool shows substantial improvements ranging from 8.2\% to 34.9\% across all models on Defects4J and from 3.3\% to 16.1\% on HumanEval-Java.
% representing 7.3\% and 6.1\% improvements over the best baseline ChatRepair. 
Additionally, \tool can also uniquely fix 7 bugs on Defects4J and 2 bugs on HumanEval-Java that all of baselines fail to resolve, proving its capability to handle challenging cases through uncertainty-guided mechanisms. 
% Additionally, \tool maintains competitive efficiency, with most configurations requiring comparable computational costs with the baselines.
\end{tcolorbox}

\subsection{RQ3: Ablation Study on \tool}
We conduct an ablation study to assess the contribution of three key components in \tool: (1) majority voting for first-token identification, (2) uncertainty-guided token-level fault localization, and (3) trace-quality measurement. 
Accordingly, we design three variants:
(1) \textbf{w/o Majority}, which removes majority voting and directly uses all initial tokens;
(2) \textbf{w/o Localize}, which replaces uncertainty-guided localization with random token selection for refinement; and
(3) \textbf{w/o Quality}, which disables trace-quality measurement by retaining all generated patches without uncertainty-based filtering.
%Specifically, we configure three variants of our tool: \tool does not utilize majority voting to handle the initial token (referred to as \textbf{w/o Majority}), the tool that does not take uncertainty-guided token-level fault localization and randomly select token for refinement (referred to as \textbf{w/o Localize}), and the tool removes trace quality measurement by always retaining a patch candidate directly without uncertainty comparison (referred to as \textbf{w/o Quality}).

Table~\ref{tab:ablation} presents the comparative results. \tool consistently outperforms all three variants, demonstrating that each component contributes meaningfully to the overall effectiveness. Removing majority voting leads to moderate performance degradation, which decreases correct fixes ranging from 2 (63 $\rightarrow$ 61) to 5 (54 $\rightarrow$ 49) cases on Defects4J and 3 (98 $\rightarrow$ 95) to 6 (110 $\rightarrow$ 104) cases on HumanEval-Java. 
However, without uncertainty-guided localization, there is the most substantial performance degradation in \tool. Correct fixes will drop ranging from 7 (53 $\rightarrow$ 46) to 13 (63 $\rightarrow$ 50) cases (13.2\% to 20.6\%) On Defects4J; 2 (95 $\rightarrow$ 93) to 13 (108 $\rightarrow$ 95) cases (2.1\% to 12.0\%) on HumanEval-Java. And removing trace quality measurement will also result in moderate losses of 3 (53 $\rightarrow$ 50) to 7 (53 $\rightarrow$ 46) correct fixes on Defects4J and 3 (95 $\rightarrow$ 92) to 6 (108 $\rightarrow$ 102) correct fixes on HumanEval-Java. The results reveals that while all three components enhance \tool's performance, uncertainty-guided token-level fault localization provides the most substantial contribution.

\begin{table*}[t]
\centering
\caption{Ablation study comparing different components of \tool}
\label{tab:ablation}
\small
\scalebox{0.85}{
\begin{tabular}{llcccccccc}
\toprule
& & \multicolumn{2}{c}{\textbf{w/o Majority}} & \multicolumn{2}{c}{\textbf{w/o Localize}} & \multicolumn{2}{c}{\textbf{w/o Quality}} & \multicolumn{2}{c}{\textbf{\tool}} \\
\cmidrule(lr){3-4} \cmidrule(lr){5-6} \cmidrule(lr){7-8} \cmidrule(lr){9-10}
\textbf{Datasets} & \textbf{Models} & $\#Plaus.$ & $\#Cor.$ & $\#Plaus.$ & $\#Cor.$ & $\#Plaus.$ & $\#Cor.$ & $\#Plaus.$ & $\#Cor.$ \\
\midrule
\multirow{5}{*}{\textbf{\textit{Defects4J}}} 
& Qwen & 78 & 61 & 68 & 50 & 76 & 58 & \textbf{80} & \textbf{63} \\
& Llama & 70 & 51 & 65 & 46 & 66 & 50 & \textbf{72} & \textbf{53} \\
& DeepSeek & 71 & 49 & 66 & 44 & 65 & 46 & \textbf{73} & \textbf{53} \\
& CodeGemma & 69 & 55 & 61 & 45 & 68 & 52 & \textbf{73} & \textbf{58} \\
& DeepSeek-V1.5 & 69 & 49 & 63 & 43 & 66 & 51 & \textbf{72} & \textbf{54} \\
\midrule
\multirow{5}{*}{\textbf{\textit{HumanEval}}} 
& Qwen & 119 & 104 & 114 & 102 & 120 & 106 & \textbf{124} & \textbf{110} \\
& Llama & 104 & 90 & 109 & 93 & 103 & 92 & \textbf{110} & \textbf{95} \\
& DeepSeek & 110 & 95 & 110 & 96 & 111 & 95 & \textbf{114} & \textbf{98} \\
& CodeGemma & 112 & 96 & 103 & 87 & 110 & 95 & \textbf{114} & \textbf{99} \\
& DeepSeek-V1.5 & 111 & 103 & 107 & 95 & 108 & 102 & \textbf{118} & \textbf{108} \\
\bottomrule
\end{tabular}
}
\end{table*}

\begin{tcolorbox}[size=title,opacityfill=0.1]
\noindent \textbf{Answer to RQ3: }
The ablation study demonstrates that all the three components contribute meaningfully to \tool's effectiveness, and uncertainty-guided fault localization has the most substantial impact, it could drop the performance by 20.6\% upon removal, validating its critical role in targeted patch refinement. The synergistic combination of these mechanisms enables \tool to achieve superior repair performance.
\end{tcolorbox}

\begin{table*}[t]
\centering
\caption{Performance comparison across different models with varying hyperparameters}
\label{tab:model_hyper}
\scalebox{0.85}{
\begin{tabular}{llccccccccc}
\toprule
\multicolumn{2}{l}{\textbf{\textit{Defects4J}}} & \multicolumn{3}{c}{$m=3$} & \multicolumn{3}{c}{$m=6$} & \multicolumn{3}{c}{$m=9$} \\
\cmidrule(lr){3-5} \cmidrule(lr){6-8} \cmidrule(lr){9-11}
 $n$ & \textbf{Models} & $\# Plaus.$ & $\# Cor.$ & $\# Gen.$ & $\# Plaus.$ & $\# Cor.$ & $\# Gen.$ & $\# Plaus.$ & $\# Cor.$ & $\# Gen.$ \\
\midrule
 \multirow{5}{*}{$n = 2$} & Qwen & \textbf{80} & \textbf{63} & 7.50 & 76 & 60 & 10.82 & 76 & 61 & 10.49 \\
& Llama & 66 & 50 & 8.47 & \textbf{72} & \textbf{53} & 9.75 & 70 & 51 & 10.51 \\
& DeepSeek & \textbf{73} & \textbf{53} & 9.82 & 72 & 52 & 7.61 & 71 & 53 & 10.52 \\
& CodeGemma & 69 & 56 & 8.97 & 63 & 50 & 11.28 & 66 & 53 & 12.63 \\
& DeepSeekV1.5 & 69 & 52 & 9.62 & 67 & 51 & 11.24 & 69 & 53 & 10.03 \\

 \multirow{5}{*}{$n = 5$} & Qwen & 77 & 61 & 8.9 & 78 & 62 & 8.35 & 68 & 56 & 8.46 \\
& Llama & 68 & 49 & 9.04 & 69 & 51 & 6.98 & 66 & 48 & 8.00 \\
& DeepSeek & 70 & 50 & 8.51 & 69 & 51 & 9.03 & 72 & 53 & 10.56 \\
& CodeGemma & 69 & 55 & 9.46 & \textbf{73} & \textbf{58} & 11.08 & 70 & 57 & 10.29 \\
& DeepSeekV1.5 & 67 & 50 & 11.03 & \textbf{72} & \textbf{54} & 11.24 & 67 & 52 & 10.70 \\
\midrule
\multicolumn{2}{l}{\textbf{\textit{HumanEval-Java}}} & \multicolumn{3}{c}{$m=3$} & \multicolumn{3}{c}{$m=6$} & \multicolumn{3}{c}{$m=9$} \\
\cmidrule(lr){3-5} \cmidrule(lr){6-8} \cmidrule(lr){9-11}
$n$ & \textbf{Models} & $\# Plaus.$ & $\# Cor.$ & $\# Gen.$ & $\# Plaus.$ & $\# Cor.$ & $\# Gen.$ & $\# Plaus.$ & $\# Cor.$ & $\# Gen.$ \\
\midrule
\multirow{5}{*}{$n = 2$} & Qwen & 123 & 108 & 4.78 & 114 & 105 & 3.82 & 116 & 107 & 3.73 \\
& Llama & 105 & 92 & 6.82 & 101 & 90 & 5.69 & 93 & 85 & 6.42 \\
& DeepSeek & 114 & 96 & 4.40 & 113 & 98 & 5.12 & 108 & 94 & 3.99 \\
& CodeGemma & 107 & 93 & 5.36 & 107 & 93 & 4.86 & 106 & 91 & 4.58 \\
& DeepSeekV1.5 & 111 & 104 & 5.31 & 120 & 108 & 7.59 & 117 & 106 & 5.20 \\

\multirow{5}{*}{$n = 5$} & Qwen & \textbf{124} & \textbf{110} & 5.42 & 117 & 106 & 4.37 & 118 & 107 & 3.49 \\
& Llama & \textbf{110} & \textbf{95} & 7.62 & 102 & 89 & 6.65 & 93 & 83 & 6.03 \\
& DeepSeek & \textbf{114} & \textbf{98} & 5.53 & 110 & 97 & 5.14 & 111 & 98 & 4.30 \\
& CodeGemma & \textbf{114} & \textbf{99} & 5.79 & 112 & 95 & 5.20 & 106 & 90 & 4.91 \\
& DeepSeekV1.5 & \textbf{118} & \textbf{108} & 5.44 & 114 & 105 & 6.19 & 114 & 104 & 5.06 \\
\bottomrule
\end{tabular}
}
\end{table*}

\subsection{RQ4: Hyperparameter Evaluation}

\textbf{Hyperparameters for Patch Generation and Refinement}. 
As specified in Algorithm~\ref{algorithm:bfs}, two key hyperparameters govern the repair process in \tool: $n$ represents the number of patches sampled from the LLM in response to each query, while $m$ denotes the number of refined patches for each suspicious token. 
% Given a fixed computational budget, the configuration of $n$ and $m$ embodies a fundamental trade-off: increasing $n$ expands the candidate pool but decrease the efforts to refine each candidate, while increasing $m$ can explore more options during refinement but decrease the number of candidate pool.
Given a fixed computational budget, the configuration of $n$ and $m$ embodies a fundamental trade-off: increasing $n$ expands the candidate pool but reduces the refinement effort allocated to each candidate, while increasing $m$ enables broader exploration during refinement but diminishes the size of the candidate pool.
To investigate the optimal balance between these two dimensions, we evaluated \tool with $n \in \{2, 5\}$ and $m \in \{3, 6, 9\}$ across all five models on both benchmarks.

Table~\ref{tab:model_hyper} presents the repair performance under different hyperparameter configurations. The results reveal that optimal parameter settings vary substantially across models and benchmarks, indicating that different models exhibit distinct characteristics in terms of initial generation quality versus remediation effectiveness. On Defects4J, models achieve their best performance under diverse configurations: Qwen performs optimally at ($n = 2$, $m = 3$) with 63 correct fixes, Llama at ($n = 2$, $m = 6$) with 53 fixes, DeepSeek at ($n = 2$, $m = 3$) with 53 fixes, CodeGemma at ($n = 5$, $m = 3$) with 58 fixes, and DeepSeek-V1.5 at ($n = 5$, $m = 6$) with 54 fixes. In contrast, on HumanEval-Java, all five models converge to the same optimal configuration ($n = 5$, $m = 3$), achieving 110, 95, 98, 99, and 108 correct fixes, respectively. A notable observation is that $m = 9$ never achieves the best performance across any model or benchmark, despite allocating the most resources to token refinement. This finding indicates that excessively aggressive remediation is counterproductive and reveals two underlying limitations. First, the effectiveness of remediation is fundamentally constrained by fault localization accuracy. When the localization component misidentifies faulty token positions, generating a larger number of refined patches (larger $m$) at incorrect positions wastes computational resources without improving repair success. Second, this phenomenon could reflect model distribution bias: if the correct replacement token has inherently low probability in the model's output distribution at the faulty position, increasing $m$ to 9 still fail to capture it. 
These observations suggest that blindly increasing $m$ within a fixed budget could encounter diminishing or even negative returns, and that balancing initial generation with moderate remediation yields superior results. 

% Based on the results, we adopt a model and benchmark-specific configuration strategy for \tool in the subsequent experiments. For the main comparison with baselines (\textbf{RQ3}) and ablation studies (\textbf{RQ4}), we select the configuration that achieves the highest number of correct fixes for each model on each benchmark, as indicated by the bold values in Table~\ref{tab:model_hyper}. This adaptive approach ensures that \tool operates at its optimal capacity for each model, providing a fair and comprehensive evaluation of the method's effectiveness across diverse model characteristics and problem domains.

\begin{tcolorbox}[size=title,opacityfill=0.1]
\noindent \textbf{Answer to RQ4: }
% The hyperparameter evaluation establishes optimal configurations in \tool. For faulty token localization, we adopt $\alpha = 0.5$ and $TopK=3$ as the default configuration across all the experiments. For hyperparameters $n$ and $m$ during patch generation, optimal configurations are model and benchmark-specific: different models achieve best performance under diverse settings. Therefore, we adopt an adaptive strategy, selecting the configuration that maximizes correct fixes for each model on each benchmark in subsequent experiments.
For patch generation and refinement, optimal ($n$, $m$) settings vary by model and benchmark, but excessively large $m$ values consistently underperform due to localization accuracy constraints and model distribution biases. Overall, moderate, balanced configurations outperform aggressive ones.
\end{tcolorbox}

\section{Related Work}
\subsection{LLM-based Program Repair}
LLM-based program repair leverages LLMs to repair program bugs automatically. The majority of LLM-based APR methods fall into two categories, distinguished by their usage approach.
\noindent \textbf{Conversation-Based Methods}. These approaches repair bugs through iterative dialogue with LLMs, incorporating program runtime feedback such as test failure information to guide the repair process. Xia et al.~\cite{xia2024automated} and Kong et al.~\cite{10.1145/3719345} pioneered conversation-based program repair systems. However, the inherent stochasticity of LLM outputs necessitates multiple repair attempts for ChatRepair~\cite{xia2024automated} and ContrastRepair~\cite{10.1145/3719345} to achieve satisfactory performance, resulting in substantial computational and financial costs. To address these limitations, Hidvégi et al.~\cite{hidvegi2024cigar} introduced CigaR, a token-efficient LLM-based APR approach designed to minimize the token overhead associated with ChatGPT. More recently, agent-based methodologies have emerged, such as RepairAgent~\cite{bouzenia2024repairagent}, which conceptualizes ChatGPT as an autonomous agent capable of planning and executing repair actions through dynamic prompting, thereby enabling more adaptive and efficient problem-solving strategies.

\noindent \textbf{Finetuning-Based Methods}.
Several fine-tuning-based approaches have been proposed for automated program repair~\cite{yuan2022circle, hao2023enhancing, wang2023rap, silva2023repairllama}. For instance, Silva et al.~\cite{silva2023repairllama} scaled fine-tuned LLMs to billion-parameter models and developed RepairLLaMA, built upon CodeLlama-7B. RepairLLaMA employs LoRA, a parameter-efficient fine-tuning technique, to train specialized repair adapters that optimize both code representation and repair capabilities.

Compared to existing studies, our work concentrates on utilizing internal signals (e.g., uncertainty) of the LLMs to conduct token-level fault localization and refine the patch from the targeted positions, thereby offering deeper insights for the development of more reliable program repair systems.

\subsection{Uncertainty of LLMs}
As large language models continue to advance across diverse application domains, quantifying uncertainty in their predictions has become increasingly critical. Uncertainty estimation provides valuable insights into model confidence, which is particularly crucial for decision-making in safety-critical domains such as medical diagnosis~\cite{fox1980evolution, simpkin2016tolerating}, where erroneous predictions can have severe consequences~\cite{alkaissi2023artificial}. Furthermore, uncertainty estimation plays a pivotal role in mitigating hallucinations in LLMs by providing a mechanism to identify when model outputs exceed the boundaries of reliable knowledge~\cite{li2023halueval}. Without robust uncertainty quantification in transformer-based systems, the trustworthiness of generated outputs remains questionable~\cite{kuhn2023semantic}.

Recent work by~\cite{spiess2024calibration} investigates the relationship between LLM-expressed confidence levels in code tokens and their corresponding accuracy in code completion tasks. Their findings reveal strong calibration between entropy-based uncertainty metrics and the correctness of generated code tokens across multiple LLMs. Specifically, the study demonstrates that tokens associated with high uncertainty exhibit a significantly higher probability of being incorrect. This observation highlights the potential of uncertainty quantification as a diagnostic tool for identifying and mitigating errors during code generation, as high-uncertainty predictions are inherently more susceptible to inaccuracies.
\section{Threats to Validity}
Manual verification of the plausible patches is a threat. A careful examination is needed to
determine whether they are semantically equivalent. To mitigate the threat, we have invested
significant labor costs to ensure the accuracy and impartiality of manual verification by inviting
three researchers in SE field to check respectively, each dedicating over 10 hours to manually
validate the patches. They then discuss the patches where validation answers were inconsistent,
ultimately reaching a consensus. The criterion for a correct patch is that it must either be identical
to the ground truth or semantically equivalent. This approach to manual verification is consistent
with methods used in several previous studies~\cite{xia2023automated, xia2024automated, ye2022selfapr}. Additionally, we have made our patches open-source for public evaluation.

Moreover, the uncertainty inherent in the experimental setup poses a potential threat to the
reproducibility of our results. Key factors of this uncertainty include the choice of temperature settings, the selection of large language models, and the non-deterministic nature
of LLM inference. For instance, minor precision errors in floating-point operations can accumulate
over time, potentially leading to significant variations in outcomes.

\vspace{-1.8mm}
\section{Conclusion}
% In this paper, we introduced ContrastRepair, a novel conversation-based APR method that utilizes
% ChatGPT for repairing bugs. By generating contrastive test pairs, ContrastRepair provides informa-
% tive feedback to ChatGPT, enabling better understanding of bug causes. We conducted extensive
% evaluations on diverse benchmark datasets, including Defects4J, QuixBugs, and HumanEval-Java,
% and compared ContrastRepair with state-of-the-art APR baselines. The results demonstrate that
% ContrastRepair achieves a new state-of-the-art in automated program repair.
In this paper, we introduced \tool, a novel uncertainty-guided APR approach that leverages token-level uncertainty to enhance LLM-based bug fixing. By incorporating context-aware uncertainty computation and fluctuation analysis, \tool precisely localizes faulty tokens within generated patches and performs targeted refinement through token-guided CoT-Decoding, enabling it to explore patch space more effectively and efficiently. We conducted extensive evaluations across two widely-used benchmarks, Defects4J 1.2 and HumanEval-Java, examining five large language models and comparing our method against three baselines. The results demonstrate that \tool achieves a new state-of-the-art in automated program repair.

% \bibliography{iclr2026_conference}
\clearpage
\bibliography{main}
\bibliographystyle{ACM-Reference-Format}

% \appendix
% \input{acmart-primary/appendix}

\end{document}